\documentclass[nofootinbib]{revtex4}
\usepackage{epsfig, graphicx}

\newcommand {\ds}{\displaystyle}
\newcommand {\fsky}{f_{\rm sky}}

\begin{document}
\title{\Large Systematic Errors in Future Weak Lensing Surveys: 
Requirements and Prospects for Self-Calibration}

\author{Dragan Huterer}
\affiliation{Kavli Institute for Cosmological Physics
and Astronomy and Astrophysics Department,
University of Chicago, Chicago, IL~~60637}

\author{Masahiro Takada}
\affiliation{Astronomical Institute, Tohoku University, Sendai
 980-8578, Japan}

\author{Gary Bernstein}
\affiliation{Department of Physics and Astronomy, University of Pennsylvania,
Philadelphia, PA 19104}

\author{Bhuvnesh Jain}
\affiliation{Department of Physics and Astronomy, University of Pennsylvania,
Philadelphia, PA 19104}

\begin{abstract}
We study the impact of systematic errors on planned weak lensing
surveys and compute the requirements on their contributions so 
that they are not a
dominant source of the cosmological parameter error budget. The generic types
of error we consider are multiplicative and additive errors in measurements
of shear, as well as photometric redshift errors. In general, more
powerful surveys have stronger systematic requirements.  For example, for a
SNAP-type survey the multiplicative error in shear needs to be smaller than
$1\%\,(\fsky/0.025)^{-1/2}$ of the mean shear in any given redshift bin, while
the centroids of photometric redshift bins need to be known to better than
$0.003\,(\fsky/0.025)^{-1/2}$. With about a factor of two degradation in
cosmological parameter errors, future surveys can enter a self-calibration
regime, where the mean systematic biases are self-consistently determined from
the survey and only higher-order moments of the systematics contribute.
Interestingly, once the power spectrum measurements are combined with the
bispectrum, the self-calibration regime in the variation of the equation of
state of dark energy $w_a$ is attained with only a 20-30\% error degradation.
\end{abstract}

\maketitle
\section{Introduction}

There has been significant recent progress in the measurements of weak
gravitational lensing by large scale structure.  Only five years after the
first detections made by several groups (Wittman et al.\ 2000, Bacon et al.\
2000, van Waerbeke et al.\ 2000, Kaiser et al.\ 2000), weak lensing already
imposes strong constraints on the matter density relative to critical
$\Omega_M$ and the amplitude of mass fluctuations $\sigma_8$ (Hoekstra, Yee \& Gladders 
2002, Jarvis et al.\ 2003, Rhodes et al.\ 2004, Heymans et al.\ 2004; for a
review see Refregier 2003), as well as the first interesting constraints on the
equation of state of dark energy (Jarvis et al.\ 2005).

The main advantage of weak lensing is that it directly probes the distribution
of matter in the universe. This makes weak lensing a powerful probe of 
cosmological parameters, including those describing dark energy (Hu \& Tegmark
1999, Huterer 2002, Hu 2003a, Heavens 2003, Refregier et al. 2003, Benabed \&
Van Waerbeke 2004, Takada \& Jain 2004, Takada \& White 2004, Song \& Knox
2004, Ishak et al.\ 2004, Ishak 2005). The weak lensing constraints are
especially effective when some redshift information is available for the source
galaxies; use of redshift tomography can improve the cosmological
constraints\ by factors of a few (Hu 1999). Furthermore, measurements
of the weak
lensing bispectrum (Takada \& Jain 2004) and purely geometrical tests (Jain \&
Taylor 2003, Bernstein \& Jain 2004, Zhang et al. 2003, Song \& Knox 2004, Hu
\& Jain 2004, Bernstein 2005) lead to significant improvements of accuracy in
measuring the cosmological parameters. When these methods are combined, weak
lensing by itself is expected to constrain the equation of state of dark energy
$w$ to a few percent, and impose interesting constraints on the time variation
of $w$. Ongoing or planned surveys, such as the Canada-France-Hawaii Telescope
Legacy Survey\footnote{http://www.cfht.hawaii.edu/Science/CFHLS}, the Dark Energy
Survey\footnote{http://cosmology.astro.uiuc.edu/DES} (DES), 
PanSTARRS\footnote{http://pan-starrs.ifa.hawaii.edu} 
and VISTA\footnote{http://www.vista.ac.uk} are expected to
significantly extend lensing measurements, while the ultimate precision will be
achieved with the SuperNova/Acceleration Probe\footnote{http://snap.lbl.gov}
(SNAP; Aldering et al.\ 2004) and the Large
Synoptic Survey Telescope\footnote{http://www.lsst.org} (LSST).

However, so far the rosy weak lensing parameter accuracy predictions that
have appeared in literature have not allowed for the presence of systematics 
(exceptions are Ishak et al.\ (2004) and Knox et al.\ (2005) who
consider a shear calibration error, and Bernstein (2005) who
does the same for the cross-correlation cosmography of weak lensing). 
This is not surprising, as we are just
starting to understand and study the full budget of systematic errors present
in weak lensing measurements. Nevertheless, some recent work has addressed
various aspects of the systematics, both experimental and theoretical, and ways
to correct for them.  For example, Vale et al.\ (2004) estimated the effects of
extinction on the extracted shear power spectrum, while Hirata \& Seljak (2003),
Hoekstra (2004) and Jarvis \& Jain (2004) considered the errors in measurements of shear.
Several studies explored the effect of theoretical uncertainties (White 2004,
Zhan \& Knox 2004, Huterer \& Takada 2004) and ways to protect against their
effects (Huterer \& White 2005). It has been pointed out that second-order
corrections in the shear predictions can be important (Schneider et al. 2002,
Hamana et al. 2002, Cooray \& Hu 2002, White 2005, Dodelson \& Zhang 2005,
Dodelson et al. 2005).  

Despite these efforts, we are at an early stage in our understanding of weak
lensing systematics.  Realistic assessments of systematic errors is likely to
impact strategies for measuring the weak lensing shear (Bernstein 2002,
Bernstein \& Jarvis 2002, Rhodes et al.\ 2004, \ Mandelbaum et al.\
2005). Eventually we would like to bring weak lensing to the same level as
cosmic microwave background anisotropies (CMB) and type Ia supernovae, where
the systematic error budget is better understood and requirements for the
control of systematic precisely outlined (e.g. Tegmark et al.\ 2000, Hu, Hedman
\& Zaldarriaga 2003, Kim et al.\ 2003, Linder \& Miquel 2004).

The purpose of this paper is to introduce the framework for the discussion of
systematic errors in weak lensing measurements and outline requirements for
several generic types of systematic error. The reason that we do {\it not}
consider specific sources of error (e.g. temporal variations in the telescope
optics or fluctuations in atmospheric seeing) is that there are many of them,
they strongly depend on a particular survey considered, and they are often
poorly known before the survey has started collecting data. Instead we argue
that, at this early stage of our understanding of weak lensing systematics, it
is more practical and useful to consider three generic types of error --
multiplicative and additive error in measurements of shear, as well as redshift
error. These generic errors are useful intermediate quantities that link actual
experimental sources of error to their impact on cosmological parameter
accuracy. In fact, realistic systematic errors for any particular experiment
can in general be converted to these three generic systematics. Given the
specifications of a particular survey, one can then estimate how much
a given systematic degrades cosmological parameters. This
can be used to optimize the design of the experiment to minimize the 
effects of systematic errors on  the accuracy of desired parameter. 
For example, accurate photometric redshift
requirements will lead to requirements on the number of filters and their
wavelength coverage. Similarly, the requirements on multiplicative and
additive errors in shear will determine how accurate the sampling of the point
spread function needs to be. 

The plan of the paper is as follows.  In \S~\ref{sec:method} we discuss the
survey specifications and cosmological parameters in this study. In
\S~\ref{sec:parameterization} we describe the parameterization of systematic
errors. In \S~\ref{sec:redshift}-\ref{sec:additive} we present the requirements
on the systematic errors for the power spectrum, while in \S~\ref{sec:bispec}
we study the requirements when both the power spectrum and the bispectrum are
used. We combine the redshift and multiplicative errors and discuss trends in
\S~\ref{sec:discussion} and conclude in \S~\ref{sec:conclude}.

\section{Methodology, Cosmological Parameters and Fiducial Surveys}
\label{sec:method}

We express the measured convergence power spectrum as
\begin{equation}
\hat{C}^{\kappa}_{ij}(\ell)=\hat{P}_{ij}^{\kappa}(\ell) + 
\delta_{ij} {\sigma_\gamma^2 \over \bar{n}_i},
\label{eq:C_obs}
\end{equation}
\noindent where $\hat{P}_{ij}^{\kappa}(\ell)$ is the measured power spectrum
with systematics (see the next Section on how it is related to the
no-systematic power spectrum $P_{ij}^{\kappa}(\ell)$), $\sigma_\gamma^2$ is the
variance of each component of the galaxy shear and $\bar{n}_i$ is the
average number of resolved galaxies in the $i$th redshift bin per steradian.  The
convergence power spectrum at a fixed multipole $\ell$ and for the $i$th and
$j$th redshift bin is given by
\begin{equation}
P_{ij}^{\kappa}(\ell) = 
\int_0^{\infty} dz \,{W_i(z)\,W_j(z) \over r(z)^2\,H(z)}\,
 P\! \left ({\ell\over r(z)}, z\right ),
\label{eq:pk_l}
\end{equation} 
\noindent where $r(z)$ is the comoving angular diameter distance and $H(z)$ is
 the Hubble parameter.  The weights $W_i$ are given by 
\begin{equation}
W_i(\chi) = {3\over 2}\,\Omega_M\, H_0^2\,g_i(\chi)\, (1+z)
\end{equation}
where $g_i(\chi) = r(\chi)\int_{\chi}^{\infty} d\chi_s n_i(\chi_s)
r(\chi_s-\chi)/r(\chi_s)$, $\chi$ is the comoving radial distance and $n_i$ is
the fraction of galaxies assigned to $i$th redshift bin.
% if $\chi_s$ falls in the distance range bounded by this bin and zero otherwise.  
We employ the redshift distribution of galaxies of the form 
\begin{equation}
n(z)\propto z^2\exp(-z/z_0)
\label{eq:nz}
\end{equation}
where $z_0$ is survey dependent and specified below.
The cosmological constraints can then be computed from the Fisher matrix
\begin{equation}
F_{ij} = \sum_{\ell} \,
\left ({\partial {\bf C}\over \partial p_i}\right )^T\,
{\bf Cov}^{-1}\,
{\partial {\bf C}\over \partial p_j},\label{eq:latter_F}
\end{equation}
\noindent where ${\bf C}$ is the column matrix of the observed power spectra and
${\bf Cov}^{-1}$ is the inverse of the covariance matrix 
between the power spectra whose elements are given by
\begin{equation}
{\rm Cov}\left [\hat{C}^{\kappa}_{ij}(\ell'), \hat{C}^{\kappa}_{kl}(\ell)\right ] = 
{\delta_{\ell \ell'}\over (2\ell+1)\,f_{\rm sky}\,\Delta \ell}\,
\left [ \hat{C}^{\kappa}_{ik}(\ell) \hat{C}^{\kappa}_{jl}(\ell) + 
  \hat{C}^{\kappa}_{il}(\ell) \hat{C}^{\kappa}_{jk}(\ell)\right ].
\label{eq:Cov}
\end{equation}
\noindent Here $\Delta \ell$ is the band width in multipole we use, and $f_{\rm
sky}$ is the fractional sky coverage of the survey. 

In addition to any nuisance parameters describing the systematics, we consider
six or seven cosmological parameters and assume a flat universe
throughout. The six standard parameters are energy density and equation of
state of dark energy $\Omega_{\rm DE}$ and $w$, spectral index $n$, matter and
baryon physical densities $\Omega_M h^2$ and $\Omega_B h^2$, and the amplitude
of mass fluctuations $\sigma_8$.  
Note that $w={\rm const}$ provides useful information about the sensitivity of an
arbitrary $w(z)$, since the best-measured mode of any $w(z)$ is about as well
measured as $w={\rm const}$ and therefore is subject to similar degradations
in the presence of the systematics. It is this particular mode, being the
most sensitive to generic systematics, that will drive the accuracy
requirements -- we explicitly illustrate this in Fig. \ref{fig:z}. In
addition to the constant $w$ case, we also consider a commonly used
two-parameter description of dark energy $w(z)=w_0+w_a z/(1+z)$ (Linder 2003)
where $w_a$ becomes the seventh cosmological parameter in the analysis.
Throughout we consider lensing tomography with 7-10 equally spaced redshift
bins (see below), and we use the lensing power spectra on scales $50\le \ell
\le 3000$.  We hold the total neutrino mass fixed at $0.1$ eV; 
the results are somewhat dependent on the fiducial mass. 
We compute the linear power spectrum using the fitting formulae of
Eisenstein \& Hu (1999).  We generalize the formulae to $w\neq -1$ by
appropriately modifying the growth function of density perturbations. To
complete the calculation of the full nonlinear power spectrum we use the
fitting formulae of Smith et al.\ (2003). 

\begin{table}
\begin{tabular}{||c|c|c|c||}
\hline\hline
                      & \rule[-2mm]{0mm}{6mm} DES  
                      & \rule[-2mm]{0mm}{6mm} SNAP    
                      & \rule[-2mm]{0mm}{6mm} LSST  \\\hline\hline
\rule[-2mm]{0mm}{6mm} Area (sq. deg.)       & 5000 & 1000    & 15000 \\\hline
\rule[-2mm]{0mm}{6mm} $n$ (gal/arcmin$^2$)  & 10   & 100     & 30    \\\hline
\rule[-2mm]{0mm}{6mm} $\sigma_\gamma$       & 0.16 & 0.22    & 0.22  \\\hline
\rule[-2mm]{0mm}{6mm} $z_{\rm peak}$        & 0.5  & 1.0     & 0.7   \\\hline\hline
\end{tabular}
\label{tab:surveys}
\caption{Fiducial sky coverage, density of source galaxies, variance of (each
component of) shear of one galaxy, and peak of the source galaxy 
redshift distribution for the three surveys considered.  }
\end{table}

The fiducial surveys, with parameters listed in Table~\ref{tab:surveys}, are:
the Dark Energy Survey; SNAP; and LSST.  Note that there is some
ambiguity in the definition of the number density of galaxies $n_g$; it is the
quantity $\sigma^2_\gamma / n_g$ that determines the shear measurement noise
level, where $\sigma_\gamma$ is the intrinsic shape noise of each galaxy.  The
surveys are assumed to have the source galaxy distribution of the form in
Eq.~(\ref{eq:nz}) which peaks at $z_{\rm peak}=2z_0$.  For the fiducial SNAP
and LSST surveys we assume tomography with 10 redshift bins equally spaced out
to $z=3$, as future photometric redshift accuracy will enable relatively fine
slicing in redshift. For the DES, we assume a more modest 7 redshift bins out
to $z=2.1$, reflecting the shallower reach of the DES while keeping the
redshift bins equally wide ($\Delta z=0.3$) as in the other two surveys.
Finally, we do not use weak lensing information beyond $\ell=3000$ in order to
avoid the effects of baryonic cooling (White 2004, Zhan \& Knox 2004, Huterer
\& Takada 2004) and non-Gaussianity (White \& Hu 2000, Cooray \& Hu 2001), both of which
contribute more significantly at smaller scales. While there may be ways to
extend the useful $\ell$-range to smaller scales without risking bias in
cosmological constraints (Huterer \& White 2005), 
%assuming $\ell_{\rm max}=3000$ is conservative and, moreover, 
extending the measurements to
$\ell_{\rm max}=10000$ would improve the marginalized errors on cosmological
parameters by only about 30\%\footnote{
On the other hand, especially for the DES, the Gaussian covariance
assumption may be somewhat optimistic for the range $1000<\ell<3000$ (e.g.\ White \& Hu 2000).}.
The parameter fiducial values and accuracies
are summarized in Table \ref{tab:errors} near the end of the article.  The
fiducial values for the parameters not listed in Table \ref{tab:errors} are
$\Omega_M h^2=0.147$, $\Omega_B h^2=0.021$, $n=1.0$, and $m_{\nu}=0.1$eV.

It is well known that measurements of the angular power spectrum of the CMB,
such as those expected by the Planck experiment, can help weak lensing constrain
the cosmological parameters. In particular, the morphology of the peaks in the
CMB angular power spectrum contains useful information on the physical matter and baryon
densities, while the locations of the peaks help constrain the dark energy
parameters.  However we checked that, when the Planck prior added, all systematics
requirements become weaker (relative to those with weak lensing alone) since
the Planck information is not degraded with systematic errors even if weak
lensing information is. In order to be conservative, we decided not
to add the Planck CMB information. Therefore, we consider the systematics
requirements in weak lensing surveys alone, and note that addition of complementary
information from other surveys typically weakens these requirements.

%\begin{table}
%\begin{tabular}{||c|c|c|c|c||}
%\hline\hline
%Parameter  & Fid.\ value
%        & \rule[-2mm]{0mm}{5mm} DES Errors 
%        & \rule[-2mm]{0mm}{5mm} SNAP Errors 
%        & \rule[-2mm]{0mm}{5mm} LSST Errors \\\hline
%$\Omega_{\rm M}$   &  0.3   & 0.008   & 0.008 & 0.003 \\\hline
%$w$                & -1.0   & 0.092   & 0.058 & 0.029  \\\hline
%$\sigma_8$         &  0.9   & 0.010   & 0.008 & 0.004  \\\hline\hline
%$w_0$              & -1.0   & 0.33   & 0.28 & 0.13  \\\hline
%$w_a$              &  0.0   & 1.41   & 0.96 & 0.49  \\\hline
%$\Omega_M h^2$     &  0.147 & 0.486   & 0.271 & 0.114  \\\hline
%$\Omega_B h^2$     &  0.021 & 0.198   & 0.102 & 0.044  \\\hline
%$n$                &  1.0   & 0.130   & 0.088 & 0.033  \\\hline
%$m_{\nu}$          &  0.1   & fixed   & fixed & fixed  \\\hline
%\hline
%\end{tabular}
%\caption{Cosmological parameters with their fiducial values and fiducial
%(no-systematics) uncertainties for the three surveys considered. Three other
%parameters ($\Omega_M h^2$, $\Omega_B h^2$ and $n$) have been marginalized
%over.  The sum of neutrino masses is held fixed at $0.1$eV. We also show the
%constraints on $(w_0, w_a)$ when these two parameters are used in place of
%constant equation of state of dark energy.  }
%\label{tab:no_sys}
%\end{table}

\section{Parameterization of the Systematics}
\label{sec:parameterization}

As mentioned above, we consider three generic sources of error. 
We believe that the parameterizations we propose, especially for the
redshift and multiplicative shear errors, are general enough to account for the
salient effects of any generic systematic. The additive shear error is more
model-dependent, and while we motivate a parameterization we believe is
reasonable at this time, further theoretical and experimental work needs to be
done to understand additive errors.

We parameterize the redshift, multiplicative shear and additive shear errors as follows.

\subsection{Redshift Errors}

Measurements of galaxy redshifts are necessary not only for the redshift
tomography -- which significantly improves the accuracy in measuring dark
energy parameters -- but also to obtain the fiducial distribution of galaxies
in redshift, $n(z)$. Therefore, understanding and correcting for the redshift
uncertainties is crucial, and comparison studies between  currently used photometric
methods, such as that initiated by Cunha et al.\ (2005), are crucial.

It is important to emphasize that statistical errors in photometric
redshifts do not contribute to the
error budget if they are well characterized. In other words, if we
know precisely the distribution of photometric redshift errors (that is, all of
its moments) at {\it each redshift}, we can use the measured photometric
distribution, $n_p(z_p)$, to recover the original spectroscopic distribution,
$n(z_s)$ very precisely. In practice we will not know the redshift
error distribution with arbitrary precision, rather
we will typically have some prior knowledge of the mean bias and scatter at each
redshift.

The quantity we consider in this paper is the uncalibrated redshift bias, that
is, the residual (after correcting for the estimated bias) offset between
the true mean redshift and the inferred mean photometric redshift at
any given $z$.
A more general description of the redshift error would include the
scatter in the redshift error at each $z$. Such an analysis has recently
been performed by Ma, Hu \& Huterer (2005) who found that, even though the
scatter is important as well, the mean bias in redshift is the dominant
source of error.

Unlike Ma, Hu \& Huterer (2005) who hold the overall distribution of galaxies
in redshift $n(z)$ fixed and only allow variations in the tomographic bin
subdivisions, we allow the redshift error to affect the overall $n(z)$ as
well. At this time it is not clear how the photometric error will affect the
source galaxy distribution $n(z)$ as it depends on how the source galaxy
distribution will be determined. We assume that the source galaxy distribution
$n(z)$ is obtained from the same photometric redshifts used to subdivide the
galaxies into redshift bins. An alternative possibility is that information
about the overall distribution of source galaxies is obtained from an
independent source (say, another survey) while the internal  photometric redshifts 
are only used to subdivide the overall distribution into redshift
bins. The two approaches, ours and that of Ma, Hu \& Huterer (2005), are
therefore complementary and both should be studied. As discussed in the
conclusions, it is reassuring that the two approaches give consistent
results.
%, largely removing  the dependence of the requirements of how
%the overall distribution of source galaxies was obtained.

We consider two alternative parameterizations of the redshift error: the
centroids of redshift bins, and the Chebyshev polynomial expansion of the mean
bias in the $z_p-z_s$ relation.  We now describe them.

\bigskip
{\it Centroids of redshift bins.}\hspace{0.2cm} We first consider the centroid
of each photometric redshift bin as a parameter.  Any scatter in galaxy
redshifts in a given tomographic bin will average out, to first order leaving
the effect of an overall bias in the centroid of this bin. [Recall, the part of
the bias that is uncalibrated and has not been subtracted out is what we consider
here.] 
As discussed by Huterer et al.\ (2004) in the context of number-count surveys,
this approach captures the salient effects of the redshift distribution
uncertainty. Note, however, that the centroid shifts do not capture the
``catastrophic'' errors where a smaller fraction of redshifts are completely
misestimated and reside in a separate island in the $z_p-z_s$ plane.

We therefore have $B$ new parameters, where $B$ is the number of redshift bins.
To compute the Fisher derivatives for these parameters, we vary each centroid
by some value $dz$, that is, we shift the whole bin by $dz$. As mentioned
above, this procedure not only allows for the fact that the tomographic bin
divisions are not perfectly measured, but it also deforms the overall
distribution of galaxies $n(z)$.

\bigskip
{\it Expansion of the redshift bias in Chebyshev polynomials.}\hspace{0.2cm}
While the required accuracy of redshift bin centroids provides useful
information, it is sometimes difficult to compare it with directly observable
quantities. In reality, an observer typically starts with measurements of
photometric redshifts which are not equal to the true, spectroscopic ones: the
quantity $z_p-z_s$ may have a nonzero value --- the bias -- and also nonzero
scatter around the biased value.  Therefore, it is sometimes more useful to
consider requirements on the accuracy in the $z_p-z_s$ relation.

Detailed forms of the bias and scatter are typically complicated and depend on
the photometric method used and how well we are able to mimic the actual
observations and correct for the biases.  We write the redshift bias as a sum
of $N_{\rm cheb}$ smooth functions -- Chebyshev polynomials centered at $z_{\rm
max}/2$ and extending from $z=0$ to $z=z_{\rm max}$, where $z_{\rm max}$ (3.0
for SNAP and LSST, 2.1 for the DES) is the extent of the distribution of
galaxies in redshift. [We briefly review the Chebyshev polynomials in the
Appendix.] The relation between the photometric and true, spectroscopic
redshifts is then

\begin{eqnarray}
z_p &=& z_s+\sum_{i=1}^{N_{\rm cheb}} g_i T_i(z_s^*), \,\,\,\,\,\,\,
   {\rm where}\label{eq:z_cheb} \label{eq:cheb_expansion}\\[0.1cm]
z_s^*&\equiv & {z_s-z_{\rm max}/2\over z_{\rm max}/2},
\end{eqnarray}

\noindent and $g_i$ are the coefficients that parameterize the bias. As with the
centroids of redshift bins, we do not model the scatter in the $z_p-z_s$
relation as one can show that the effect of uncalibrated bias is dominant (Ma,
Hu \& Huterer, 2005). The effect of imperfect redshift measurements is to shift
the distribution of galaxies away from the true distribution $n(z_s)$ to a
biased one $n_p(z_p)$. The biased distribution of galaxies then propagates to
bias the cosmological parameters. The photometric distribution can then simply
be obtained from the true distribution as (e.g.\ Padmanabhan et al.\ 2005)

\begin{equation}
n_p(z_p)=\int_0^\infty n(z) \Delta(z_p-z, z) dz
\label{eq:np}
\end{equation}

\noindent where $\Delta(z_p-z, z)$ is the probability that the galaxy at redshift $z$
is measured to be at redshift $z_p$. Since we are not modeling  the scatter in
the $z_p-z_s$ relation, the probability is a delta function

\begin{equation}
\Delta(z_p-z, z) = \delta\left (z_p-z_s-\sum_{i=1}^{N_{\rm cheb}} g_i T_i(z_s^*)\right ).
\label{eq:Delta}
\end{equation}

\noindent Since we will include the $g_i$ as additional parameters, with
fiducial values $g_i=0$, we shall need to take derivatives with one nonzero
$g_i$ at a time and therefore we can assume $\Delta(z_p-z, z_p) =
\delta(z_p-z_s-g_i T_i(z_s^*))$ for a single $i$.  Using the fact that
$\delta(F(x))=\sum 1/|F'(x_0)|\, \delta(x-x_0)$ for a function $F(z)$, where
the sum runs over the roots of the equation $F(x)=0$, and further using a recursive
formula for the derivative of the Chebyshev polynomial (Arfken 2000,
\S\ 13.3), we get

\begin{equation}
n_p(z_p) = \sum_a 
{n(z_a)\over \left | 1+ {\ds 2g_i\over \ds z_{\rm max}[1-(z_a^*)^2]} 
  \ds\left [ -iz_aT_i(z_a^*)+iT_{i-1}(z^*_a)\right ] \ds\right | }
\label{eq:np_final}
\end{equation}

\noindent where the sum runs over all roots of the equation $z+g_i
T_i(z^*)-z_p=0$, with (recall) $z^*_a\equiv (z_a-z_{\rm max}/2)/ (z_{\rm
max}/2)$.  This is the expression for the perturbed distribution of galaxies
due to a single perturbation mode.\footnote{Note that, for $g_i\ll 1$,
Eq.~(\ref{eq:np_final}) simplifies to $n_p(z_p)=n(z_p-g_i T_i(z_p^*))$} We use
it, together with the original distribution $n(z)$, to compute the perturbed
and unperturbed convergence power spectra and thus take the derivative
with respect to $g_i$.

\begin{figure}
\psfig{file=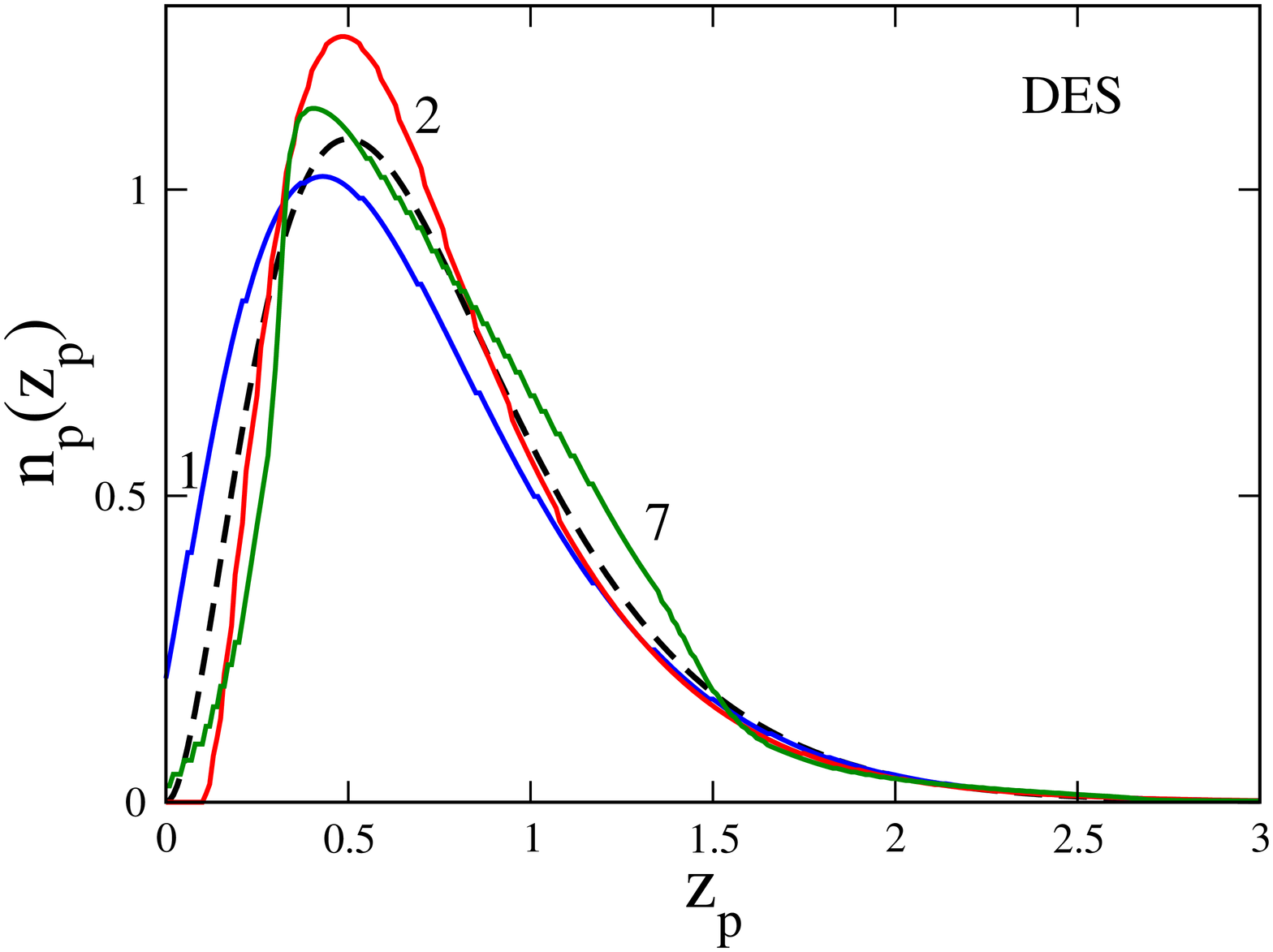, width=2.8in, height=3.5in, angle=-90}\hspace{-0.2cm}
\psfig{file=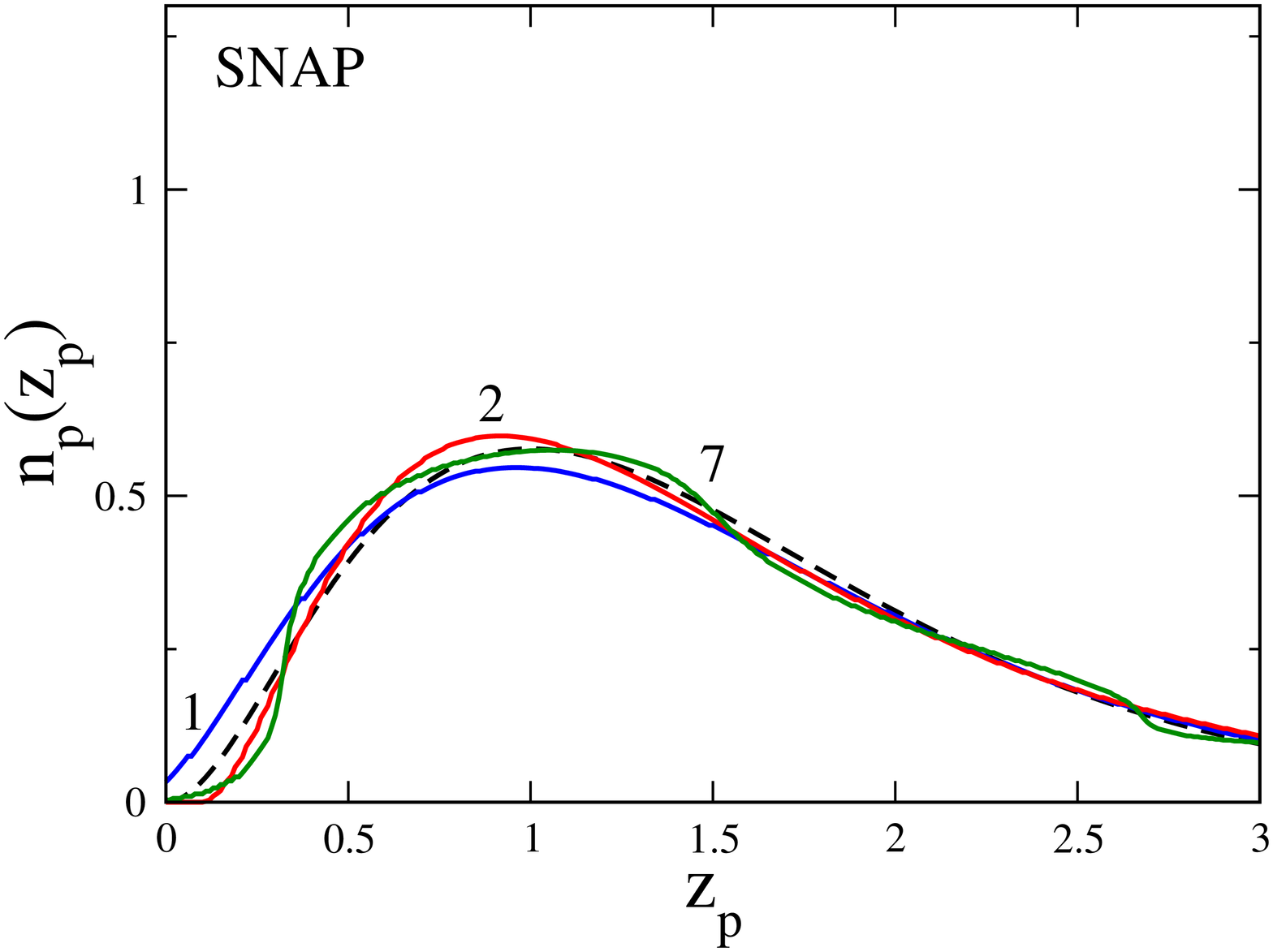, width=2.8in, height=3.5in, angle=-90}
\caption{Select three modes (first, second and seventh) of perturbation to the
distribution of galaxies $n(z)$ for the DES (left) and SNAP (right), shown
together with the original unperturbed distribution (dashed curve in each
panel).  These three modes correspond to the modes of perturbation in the
$z_p-z_s$ relation shown in Fig.~\ref{fig:deltaz_vs_zs} in the Appendix.  Note
that SNAP's fiducial $n(z)$ is broader, making the resulting wiggles in
$n_p(z_p)$ less pronounced and the weak lensing measurements therefore less
susceptible to the redshift biases.  Note also that the allowed perturbations
to $n(z)$ are not only located near the peak of the distribution, but can also
have significant wiggles near the tails of the distribution. }
\label{fig:nz_pert}
\end{figure}

In Fig.~\ref{fig:nz_pert} we show the photometric galaxy distributions
$n_p(z_p)$ for the select three Chebyshev modes (first, second and seventh) of
perturbation to the distribution of galaxies $n(z)$ corresponding to
Fig.~\ref{fig:deltaz_vs_zs}, together with the original unperturbed
distribution.  The distributions are shown for the DES and SNAP.  Since the
incorrect assignment of photometric redshifts only redistributes the galaxies,
we always impose the requirement $\int_0^\infty n_p(z_p) dz_p=1$ regardless of
the perturbation.  Note that SNAP's fiducial $n(z)$ is broader, making the
resulting wiggles in $n_p(z_p)$ less pronounced and helping the weak lensing
measurements be less susceptible to the redshift biases. (As shown in
\S~\ref{sec:redshift_centroids}, this effect is counteracted by the smaller
fiducial errors in the SNAP survey which lead to more susceptibility to biases.)

\subsection{Multiplicative errors}
\label{sec:mult_define}

The multiplicative error in measuring shear can be generated by a variety of
sources.  For example, a circular PSF of finite size is convolved with the true
image of the galaxy to produce the observed image, and in the process it
introduces a multiplicative error. Let $\hat{\gamma}(z_s, {\bf n})$ and
$\gamma(z_s, {\bf n})$ be the estimated and true shear of a galaxy at some true
(spectroscopic) redshift $z_s$ and direction ${\bf n}$. Then the general
multiplicative factor $f_i({\mathbf \theta})$ acts as

\begin{equation}
\hat{\gamma}(z_s, {\bf n}) = \gamma\left (z_s, {\bf n}\right )\times 
    [1+f_i(z_s, {\bf n})] 
\end{equation}

\noindent where $f_i$ is the multiplicative error in shear which is both direction and
time dependent. We can write the error as a sum of its mean (average over all
directions and redshifts in that tomographic bin) and a component with zero mean,

\begin{equation}
f(z_{s,i}, {\bf n})=f_i+r_i({\bf n})
\end{equation} 

\noindent where \noindent $\langle f(z_{s,i}, {\bf n})\rangle=f_i$ and $\langle r_i({\bf
n})\rangle=0$ and the averages are taken over angle and over redshift 
within the $i$th redshift bin. Then we can write

\begin{eqnarray}
  \langle \hat{\gamma}(z_{s,i}, {\bf n})\,
  \hat{\gamma}(z_{s,j}, {\bf n} +d{\bf n})\rangle
  &=& \langle \gamma(z_{s,i}, {\bf n})\,\gamma(z_{s,j}, {\bf n} +d{\bf n})
  \,(1+f(z_{s,i}, {\bf n}))\,(1+f(z_{s,j}, {\bf n}))\rangle \\[0.1cm]
%  &\approx & \langle \gamma(z_{s,i}, {\bf n})
%                     \gamma(z_{s,j}, {\bf n} +d{\bf n})\rangle
%           + \langle \gamma(z_{s,i}, {\bf n})
%                     \gamma(z_{s,j}, {\bf n} +d{\bf n})\rangle
%                     (f_i+f_j)\\[0.1cm]
     & \simeq &    \langle \gamma(z_{s,i}, {\bf n})
                     \gamma(z_{s,j}, {\bf n} +d{\bf n})\rangle
                     (1+f_i+f_j)
\end{eqnarray}
\noindent where we dropped terms of order $\langle \gamma\, \gamma\,f_i
f_j\rangle$. Since the random component of the
multiplicative error is uncorrelated with shear, all terms of the form
$\langle \gamma\, \gamma\, r\rangle$ are zero. Finally, terms of the form
$\langle \gamma\, \gamma\, r\,r\rangle$ are taken to be small, thus
requiring that $\langle r_i({\bf n})\,r_j({\bf n + dn})\rangle$ is
smaller than $f_i$ and $f_j$ (Guzik and Bernstein, 2005).
Therefore
\begin{equation}
\hat{P}^{\kappa}_{ij} (\ell)
=    P^{\kappa}_{ij} (\ell) \times\left [1+f_i+f_j\right ] 
\label{eq:mult}
\end{equation}

\noindent where again we emphasize that $f_i$ is the irreducible part of the
multiplicative error in $i$th redshift bin (i.e.\ its average over all
directions {\it and}  the $i$th redshift slab).  It is clear that 
multiplicative errors are potentially dangerous, since they lead to 
an error that goes as
$f_i$ and not $f_i^2$.  Finally, note that shear calibration is
likely to depend on the size of the galaxy, so that the mean error $f_i$ is the
error averaged over all galaxy sizes in bin $i$.

\subsection{Additive errors}\label{sec:add_define}

Additive error in shear is generated, for example, by   the
anisotropy of the PSF. 
%, which adds a fixed amount of shear to all
%galaxies at some location. 
We define the additive error via

\begin{equation}
\hat{\gamma}(z_s, {\bf n}) = \gamma (z_s, {\bf n})+\gamma_{\rm add} (z_s, {\bf n})
\end{equation}
Let us assume that the additive error is uncorrelated with
the {\it true} shears so that the term $\langle \gamma({\bf
n})\gamma_{\rm add}({\bf n})\rangle$ is zero. Furthermore, let the Legendre
transform of the term $\langle \gamma_{\rm add}({\bf n})\gamma_{\rm
add}({\bf n})\rangle$ be $P^{\kappa}_{\rm add}(\ell)$. Then we can write
\begin{equation}
\hat{P}^{\kappa}_{ij} (\ell)
=    P^{\kappa}_{ij}(\ell) + P^{\kappa}_{\rm add}(\ell).
\end{equation}

We now have to specify the additive systematic power $P^{\kappa}_{\rm
add}(\ell)$.  Consider some known sources of additive error: for example, the
additive error induced due to a non-circular PSF is roughly $R\,e_{\rm PSF}$
where $e_{\rm PSF}$ is the ellipticity of the PSF and $R\equiv (s_{\rm
PSF}/s_{\rm gal})^2$ is the ratio of squares of the PSF and galaxy size
(E. Sheldon, private communication); galaxies that are smaller therefore have a
larger additive error. Of course, the observer needs to correct the overall
trend in the error, leaving only a smaller residual, and the impact of this
residual is what we are interested in. It should be clear from this discussion
that the additive error in shear can be calibrated by a part that depends on
the average size of a galaxy at a given redshift multiplied by a random
component that depends on the part of the sky observed.  Therefore, we write
the additive shear as $\gamma_{\rm add}(z_{s,i}, {\bf n})=b_ir({\bf n})$ where
$b_i$ is the characteristic additive shear amplitude in the $i$th redshift bin
and $r({\bf n})$ is a random fluctuation.

%Consider, for example, scanning strategy of SNAP telescope that traces out
%stripes on the sky each consecutive period. The telescope optics can vary on
%short timescales (seconds or minutes) while averaging out over a period of
%days.  In that case, we would expect the error to be larger on small angular
%scales (high $\ell$) while averaging over large angular scales (Stabenau et
%al., 2005).  We choose to model the angular part in multipole space as a power
%law in $\ell$; the expectation is that the correlation will increase toward
%small scales.  therefore, the additive error has power even in
%$P^{\kappa}_{ij}$ where $i$ and $j$ are distinct redshift bins. 
In multipole space, the two point function can then be expressed as $b_i b_j$
times the angular part which only depends on $|{\bf n}_i -{\bf n}_j|$. Note
that the additive error for two galaxies at different redshifts (i.e.\ when
$i\neq j$) is not zero, although it may in principle be suppressed relative to
the additive error for the auto-power spectra (when $i=j$). Motivated by such
considerations, we assume the additive systematic in multipole space of the
form

\begin{equation}
P_{{\rm add}, ij}^{\kappa}(\ell)=\rho \,  b_ib_j 
\left ({\ell\over \ell_*}\right )^{\alpha}
\label{eq:add}
\end{equation}

\noindent where the correlation coefficient $\rho$ describes correlations
between bins.  The coefficient $\rho$ is always set to unity for $i=j$, and for
$i\ne j$ it is fixed to some fiducial value (not
taken as a parameter in the Fisher matrix).
While the discussion above would imply that $\rho=1$ for all $i$ and $j$, we
allow for the possibility that the additive errors are not perfectly
correlated across redshift bins.
In the extreme case of uncorrelated additive errors in different $z$
bins, the cross-power spectra are not 
affected, $\rho=\delta_{ij}$. We shall see that the results are weakly
dependent on the fiducial value of $\rho$ unless if $\rho= 1$ identically.  

For the multipole dependence of $P^\kappa_{\rm add}$ we assume a
power-law form, and marginalize over the index $\alpha$.  
We choose $\ell_*=1000$, near the ``sweet spot'' of weak lensing
surveys, but note that this choice is arbitrary and made solely for
our convenience since $\ell_*$ is degenerate with the parameters $b_i$. Our
{\it a priori} expectation for the value of $\alpha$ is uncertain, and we try
several possibilities but find (as discussed later) that the results are
insensitive to the value of $\alpha$.  We therefore have a total of $B+1$
nuisance parameters for the additive error ($B$ parameters $b_i$, plus
$\alpha$).

In practice there is no particular reason why an additive systematic would be
expected to conform to a power law, so in the future we should consider an
additive systematic with more freedom in its spatial behavior.  The current
treatment is optimistic: fixing the systematic to a power law gives it an
identifying characteristic that allows us to distinguish it from cosmological
signals. Without such a characteristic, the systematic could be much more
damaging.

\subsection{Putting the systematics together}
\label{sec:all_sys}

With the systematic errors we consider, the resulting theoretical convergence
power spectrum is

\begin{equation}
\hat{P}^{\kappa}_{ij} (\ell)
\equiv \hat{P}^{\kappa}(\ell, z_s^{(i)}\,,\, z_s^{(j)})
=    P^{\kappa}
\left (\ell, z_s^{(i)}+\delta z_p^{(i)}\,,\, z_s^{(j)}+\delta z_p^{(j)}\right )
\times\left [1+f_i+f_j \right ] + \rho\, b_ib_j  
\left ({\ell\over \ell_*}\right )^{\alpha}.
\end{equation}

The derivatives with respect to the multiplicative and additive parameters are
trivial since they enter as linear (or power-law) coefficients, while the
derivatives with respect to the redshift parameters $\delta z_p^{(k)}$ (really,
parameters $g_i$ defined in Eq.~(\ref{eq:cheb_expansion})) need to be taken
numerically.

\section{Results: redshift systematics}
\label{sec:redshift}

\subsection{Centroids of redshift bins}
\label{sec:redshift_centroids}

We have a total of $N+B$ parameters, and we marginalize over the $B$ redshift
bin centroids by giving them identical Gaussian priors. While the
actual redshift accuracy may be better in some redshift ranges than in others,
the required accuracy that we obtain in our approach will pertain to the redshift
bins where observations are most sensitive (i.e. at intermediate redshifts $z\sim 0.5-1$).

\begin{figure}[!t] 
\psfig{file=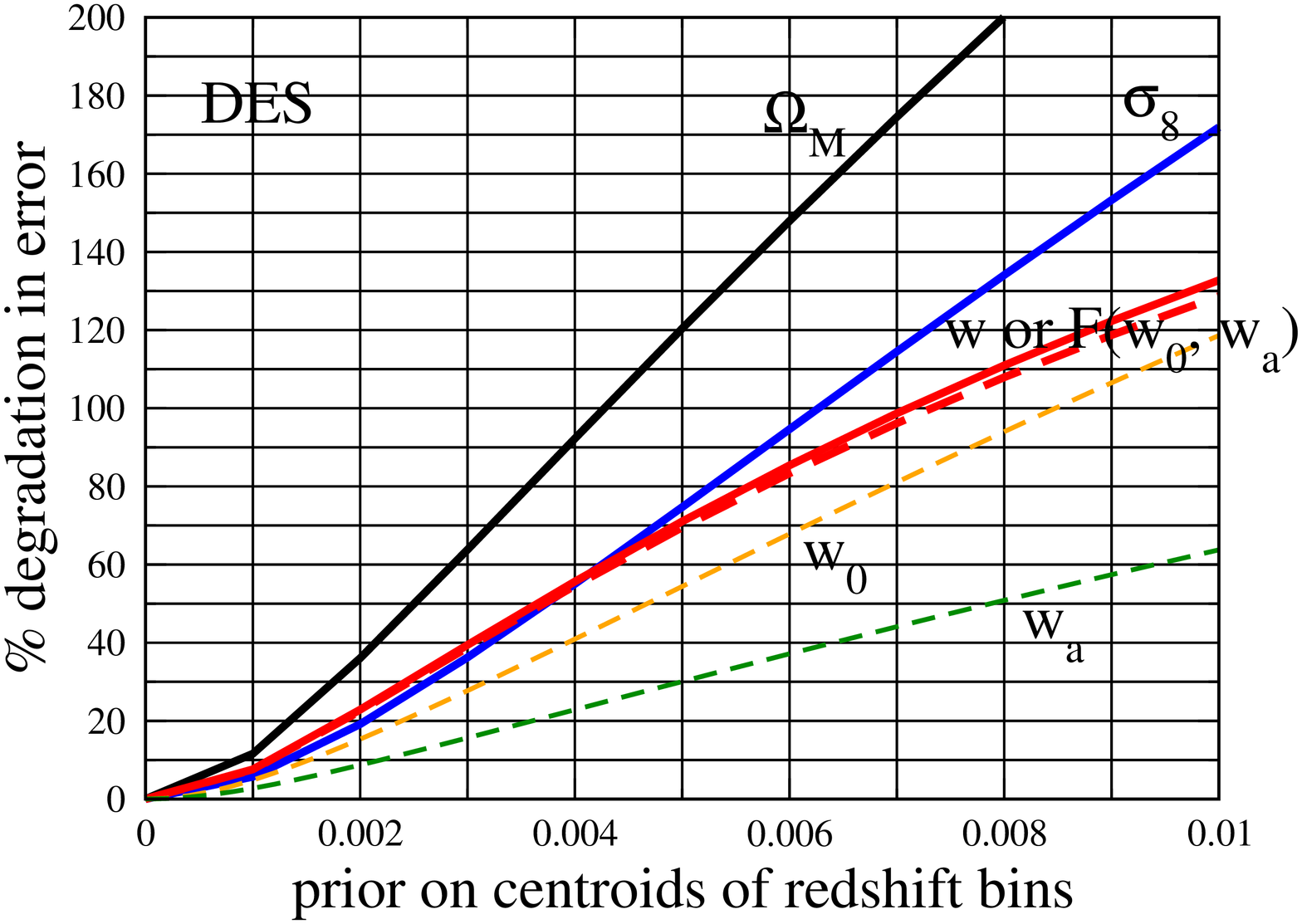, width=2.8in, height=3.5in, angle=-90}\hspace{-0.2cm}
\psfig{file=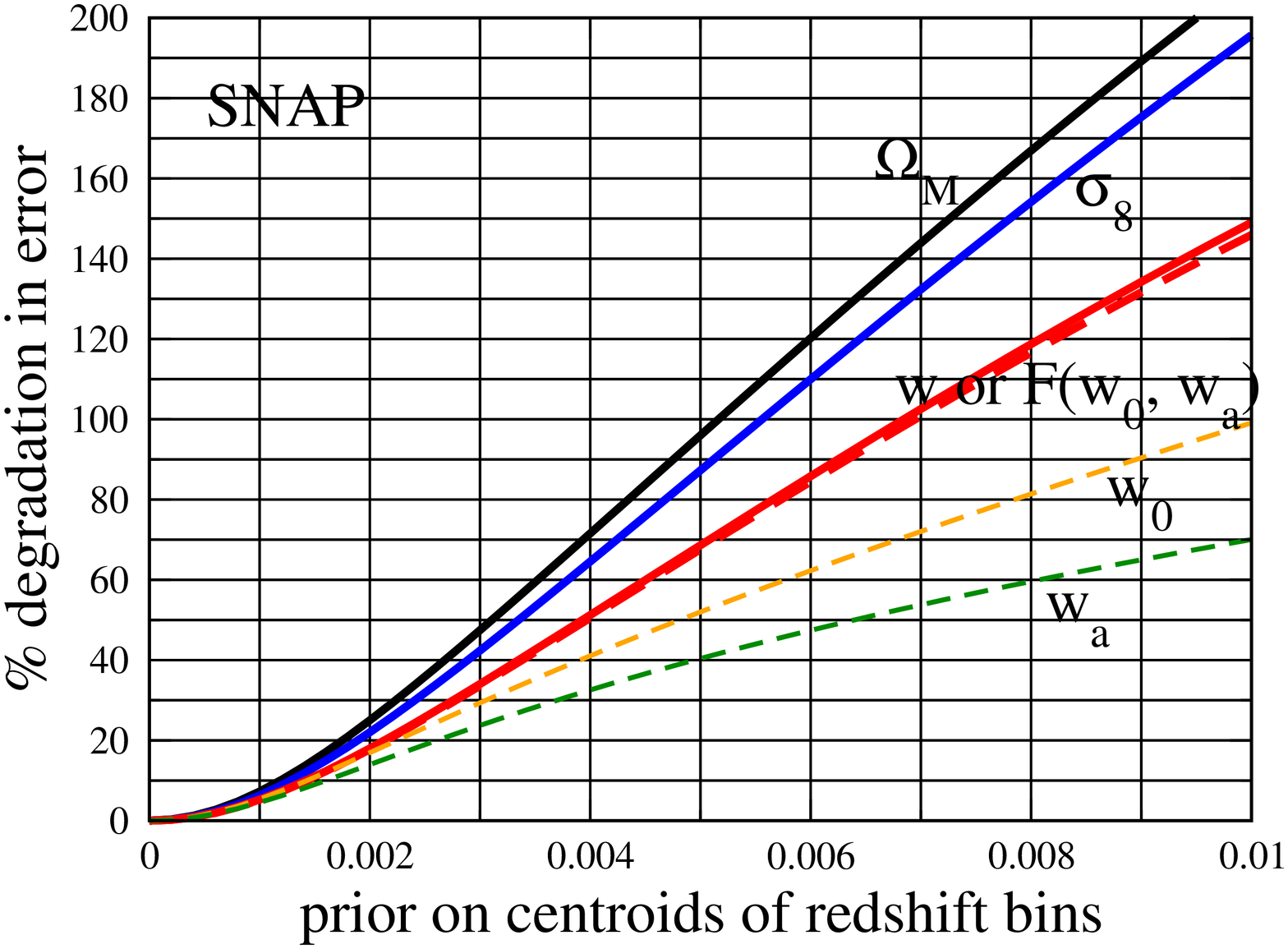, width=2.8in, height=3.5in, angle=-90}\\[-0.1cm]
\psfig{file=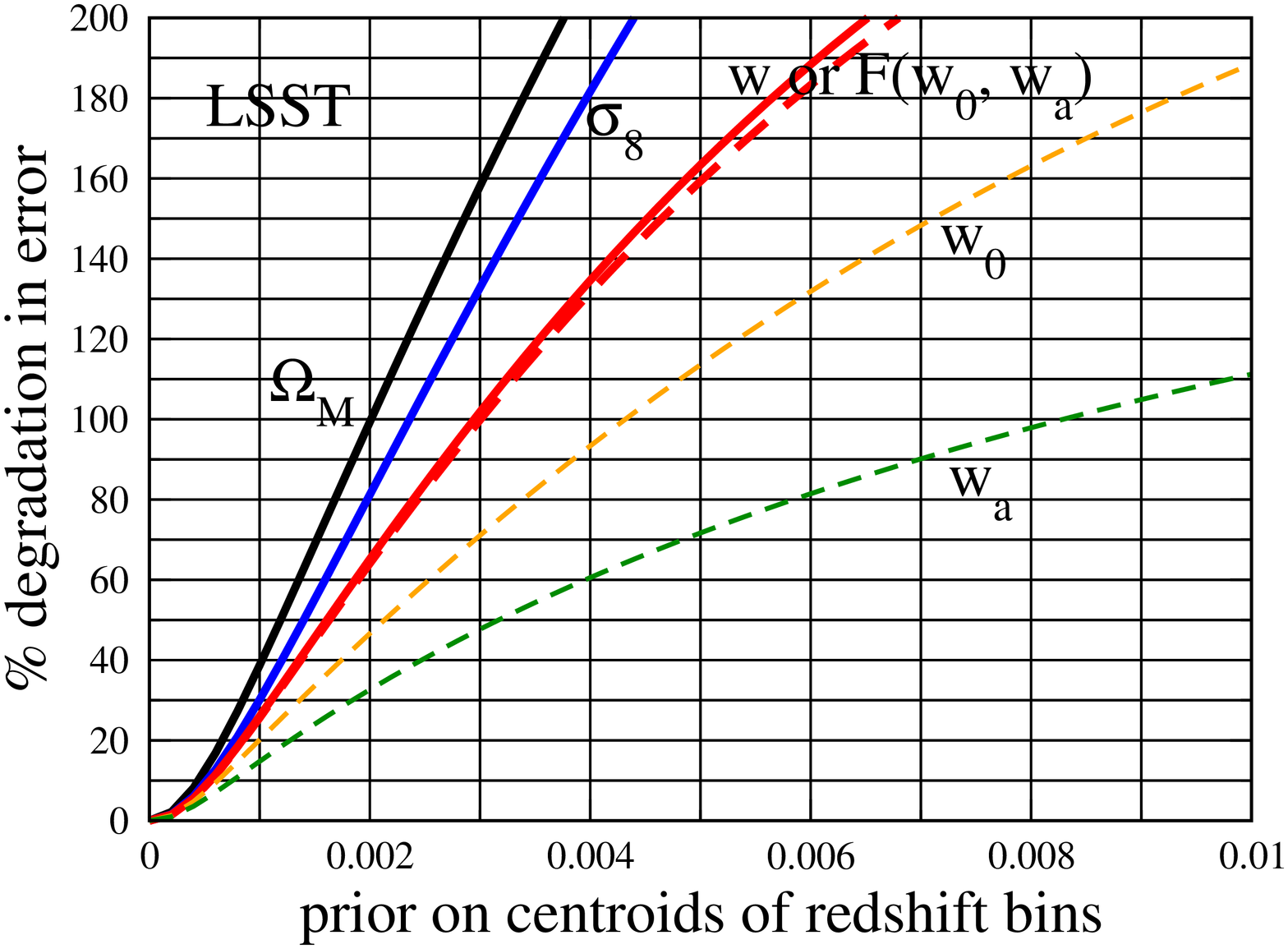, width=2.8in, height=3.5in, angle=-90}
\caption{Degradation in the cosmological parameter accuracies as a function of
our prior knowledge of $\delta z \equiv z_p - z_s$. We assume equal Gaussian
priors to each redshift bin centroid, shown on the x-axis.  For example, to
have less than $\sim 50\%$ degradation in $\Omega_M$, $\sigma_8$ or $w$, we
need to control the redshift bias to about $0.003\,(\fsky/f_{\rm sky,
fid})^{-1/2}$ or better for the DES and SNAP, and to about $0.0015\,(\fsky/f_{\rm
sky, fid})^{-1/2}$ or better for the LSST. For the varying equation of state
parameterization, the requirements for the best measured
combination of $w_0$ and $w_a$ ($F(w_0, w_a)$) are identical to those for $w={\rm const}$, 
while the requirements on $w_0$ and $w_a$ individually are somewhat less stringent.}
\label{fig:z}
\end{figure}

Figure~\ref{fig:z} shows the degradation in $\Omega_M$, $\sigma_8$ and $w={\rm
const}$, and also $w_0$ and $w_a$, accuracies as a function of our prior
knowledge of the redshift bin centroids. Here and in the subsequent two
Figures, the plots are shown separately for each of the three surveys (DES,
SNAP and LSST). Since we are using the Fisher matrix formalism without external
priors on cosmological parameters, the cosmological parameter degradations (as
well as accuracies) clearly remain unchanged for an arbitrary $\fsky$ if we
also scale the priors on the nuisance parameters by $\fsky^{-1/2}$.  In these
and subsequent plots we show the degradations for fiducial values of the sky
coverage for each survey $f_{\rm sky, fid}$, where $f_{\rm sky, fid}=f_{5000}$,
$f_{1000}$ and $f_{15000}$ for DES, SNAP and LSST respectively, while
indicating that the requirements scale as $\,(\fsky/f_{\rm sky, fid})^{-1/2}$.

We generally find that the degradations in different cosmological parameters
are comparable.  To have less than $\sim 50\%$ degradation, for example, we
need to control the redshift centroid bias to about 0.003$\,(\fsky/f_{\rm sky,
fid})^{-1/2}$ for the DES and SNAP and to about 0.0015$\,(\fsky/f_{\rm sky,
fid})^{-1/2}$ for the LSST. The current statistical accuracy in {\it
individual} galaxy redshifts is of order 0.02-0.05. Averaging the large number
of galaxies in a given redshift bin, it may be that we are already close to the
aforementioned requirement on the centroid bias, but this will of course depend
on the control of systematics in the photometric procedure. 

As mentioned in the Introduction, the accurately determined combination of $w_0$
and $w_a$, $F(w_0, w_a)$, is degraded in nearly the same way as constant $w$;
this is illustrated by a thick dashed line in each panel of
Fig.~\ref{fig:z}. In the rest of the paper we do not repeat plotting the curves
corresponding to $F(w_0, w_a)$ which nearly overlap those corresponding to
$w={\rm const}$. Finally, we show the degradations for $w_0$ and $w_a$
separately.  Since these two parameters are determined to a substantially lower
accuracy than $w={\rm const}$ (see Table \ref{tab:errors}), it is not
surprising that the degradations in these two are substantially smaller.

\subsection{Chebyshev expansion in $z_p-z_s$}

\begin{figure}[!h]
\psfig{file=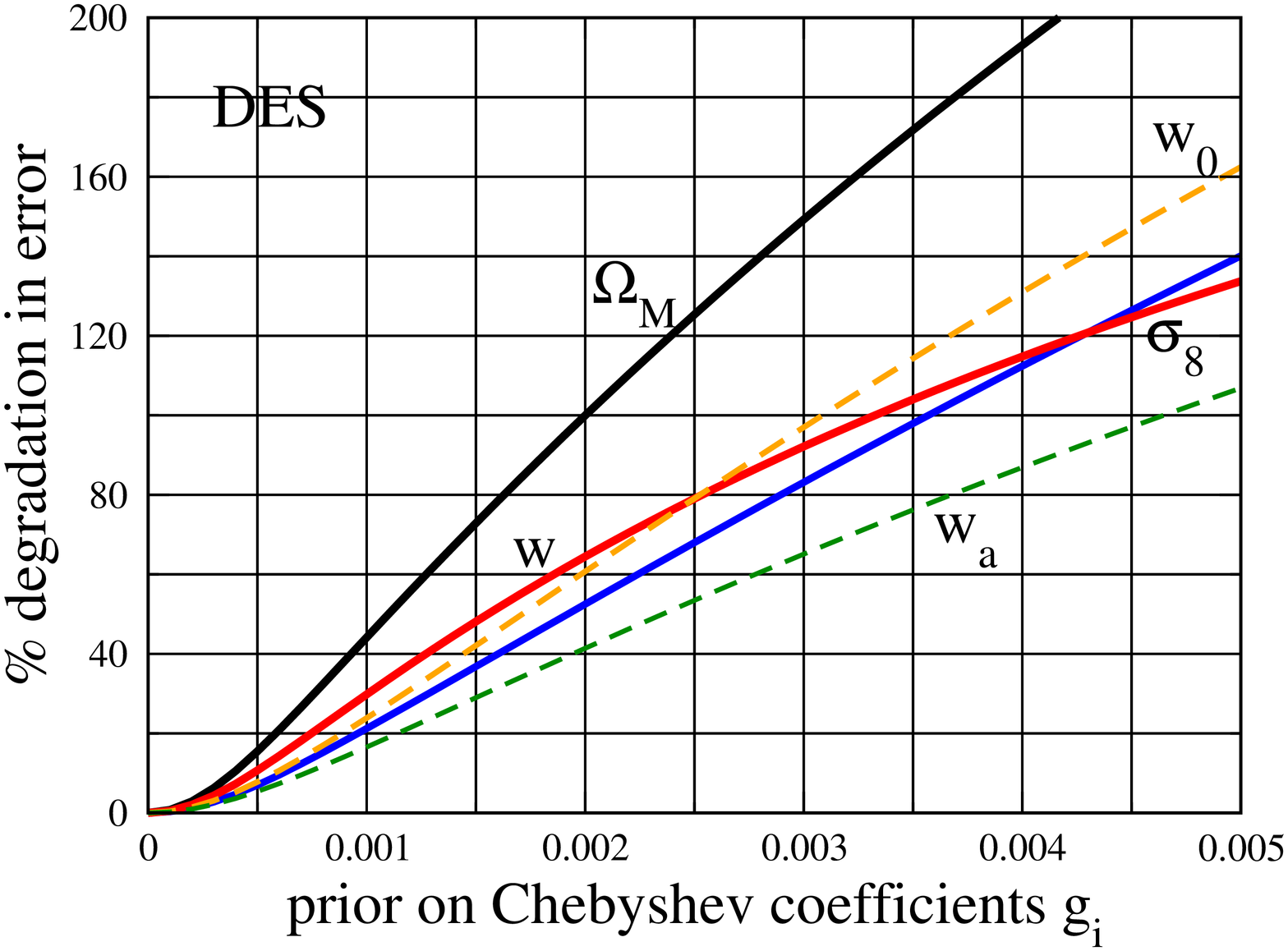, width=2.8in, height=3.5in, angle=-90}\hspace{-0.2cm}
\psfig{file=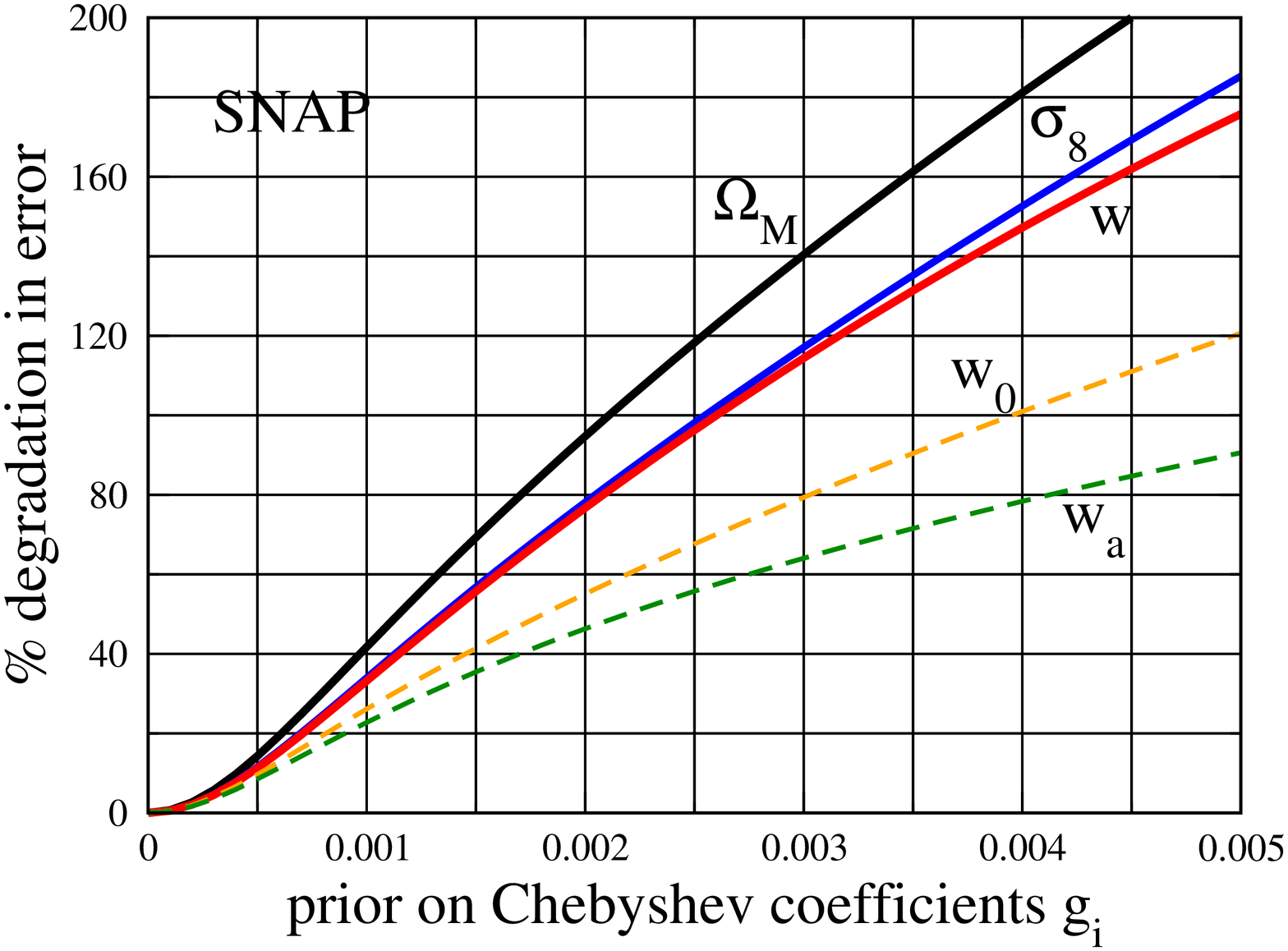, width=2.8in, height=3.5in, angle=-90}\\[-0.1cm]
\psfig{file=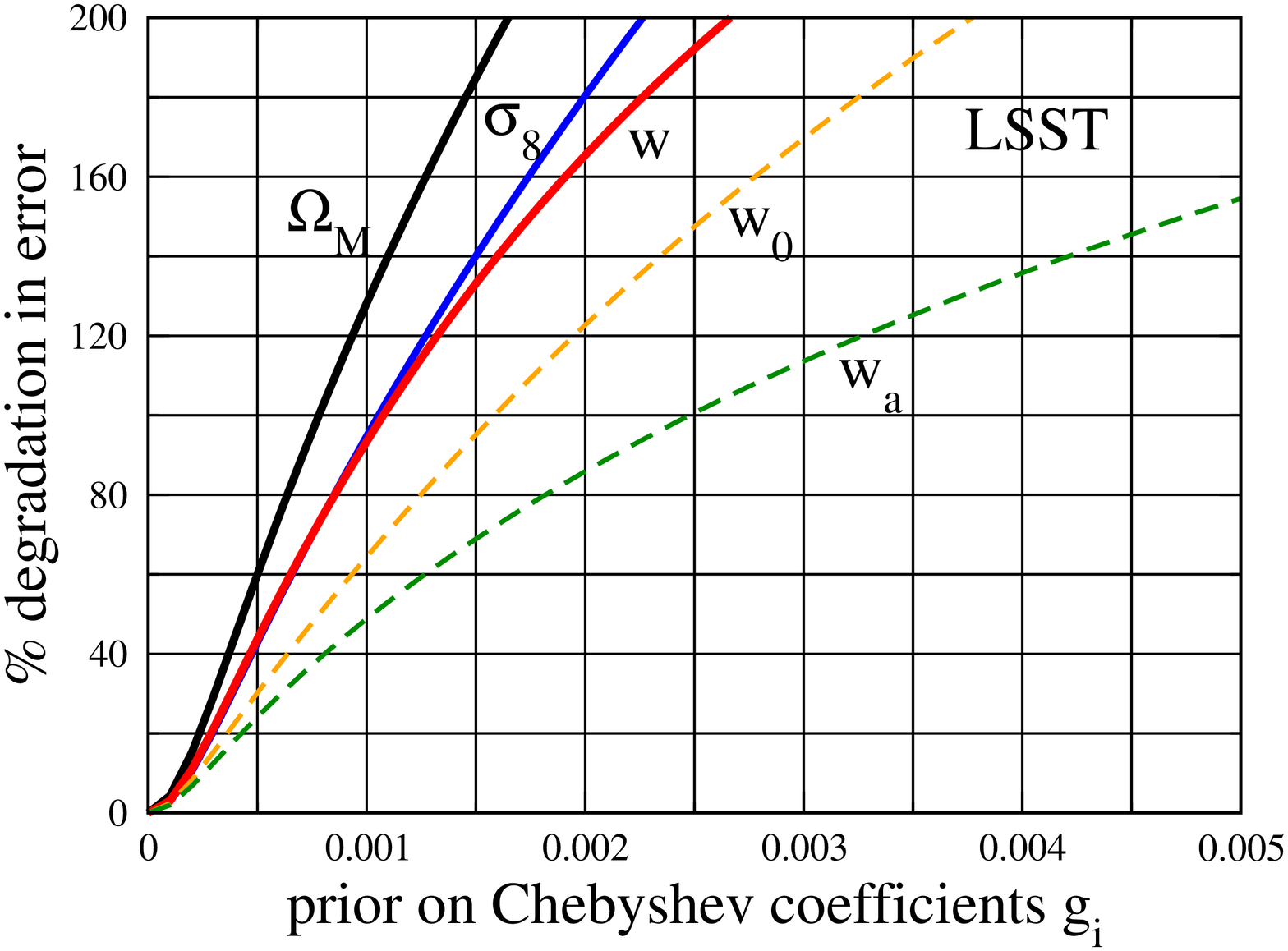, width=2.8in, height=3.5in, angle=-90}
\caption{Degradation in marginalized errors in $\Omega_M$, $\sigma_8$ and
$w={\rm const}$, as well as $w_0$ and $w_a$, as a function of our prior
knowledge of the redshift bias coefficients $f_i$ for the DES, SNAP and
LSST. We use $N_{\rm cheb}=30$ parameters $f_i$ that describe the bias in
redshift and give equal prior to each of them, shown on the x-axis. For the DES
and SNAP, knowledge of $f_i$ to better than $0.001\,(\fsky/f_{\rm sky,
fid})^{-1/2}$, corresponding to redshift bias of $|z_p-z_s|\lesssim
0.001\,(\fsky/f_{\rm sky, fid})^{-1/2}$ for each Chebyshev mode, is desired as
it leads to error degradations of about 50\% or less.  For LSST the requirement
is about a factor of two stronger.  These results are corroborated by computing
the bias in cosmological parameters as discussed in the text. }
\label{fig:ncheb_req}
\end{figure}

Figure \ref{fig:ncheb_req} shows the degradation in the $\Omega_M$, $\sigma_8$
and $w$, and also $w_0$ and $w_a$, as we marginalize over a large number ($N_{\rm
cheb}=30$) of Chebyshev coefficients $f_i$ each with the given prior. We
checked that, when only $N\lesssim 5$ modes are allowed to vary, all parameters
$f_i$ can be self-consistently solved from the survey -- this is an example of
self-calibration. However, with a larger number of Chebyshev modes the
self-calibration regime is lost.  Since we are interested in the most general
form of the redshift bias, this is the regime we would like to explore.  We
have also checked that, as we let the number of coefficients increase further,
the errors in cosmological parameters do not increase indefinitely, as the
rapid fluctuations in the $z_p-z_s$ are not degenerate with cosmology. In fact
we found that the degradations have asymptoted to their final values once we
include 20-30 Chebyshev polynomials; therefore, fluctuations in redshift on
scales smaller than $\Delta z\sim 0.1$ are not degenerate with cosmological
parameters. The choice of $N_{\rm cheb}=30$ coefficients is therefore
conservative.

Figure \ref{fig:ncheb_req} shows that the photometric redshift bias in each
Chebyshev mode, for the DES and SNAP, should be controlled to about
$0.001\,(\fsky/f_{\rm sky, fid})^{-1/2}$ or less.
For LSST, the requirement is more severe and we could tolerate biases no larger
than about $0.0005\,(\fsky/f_{\rm sky, fid})^{-1/2}$. These are fairly
stringent requirements, in rough agreement with results found by Ma, Hu \&
Huterer (2005).

It might seem a bit surprising that the SNAP and DES requirements are
comparable, given that the fiducial SNAP survey is more powerful than the DES
(see Table \ref{tab:errors}) and hence would need better control of the
systematics. However, there is another effect at play here: a narrow
distribution of galaxies in redshift, such as that from the DES, is more
strongly subjected to fluctuations in $z_p-z_s$ than a wide distribution, such
as that from SNAP; see Fig.~\ref{fig:nz_pert}.  Therefore, surveys with wide
leverage in redshift have an advantage in beating down the effect of redshift
error, just as in the case of cluster count surveys (Huterer et al.\ 2004).

\bigskip

We now confirm the results by computing the bias in  the cosmological parameters
due to the bias in one of the redshift bias coefficients, $dg_i$. The bias in the
cosmological parameter  $p_{\alpha}$ can  be computed as

\begin{equation}
\delta p_{\alpha}  =  F_{\alpha \beta}^{-1}\, \sum_{\ell}
\left [C_i^{\kappa}(\ell)-\bar{C}_i^{\kappa}(\ell)\right ]
 {\rm Cov}^{-1}\left [\bar C_i^{\kappa}(\ell), 
  \bar C_j^{\kappa}(\ell)\right ]\,
   {\partial \bar C_j(\ell)^{\kappa} \over \partial p_{\beta}}
\label{eq:bias}
\end{equation}

\noindent where the summations over $\beta$, $i$ and $j$ were implicitly
assumed and the covariance of the cross power spectra is given in
Eq.~(\ref{eq:Cov}).  The source of the bias in the observed shear covariance,
$C_i^{\kappa}(\ell)-\bar{C}_i^{\kappa}(\ell)$, is assumed to be the excursion
of $dg_m=0.001$ in the $m$th coefficient of the Chebyshev expansion, others
being held to their fiducial values of zero. We then compare this error to the
1-$\sigma$ marginalized error on the parameter $p_{\alpha}$.  For a range of
$1\leq m\leq 30$ we find that the biases in the cosmological parameters are
entirely consistent with statistical degradations shown in
Fig.~\ref{fig:ncheb_req}.  We also find that the degradations due to higher
modes ($m\gtrsim 10$-$20$) are progressively suppressed, illustrating again that
smooth biases in redshift are most important source of degeneracy with
cosmological parameters, and need the most attention when calibrating the
photometric redshifts.

Finally, let us note that in this analysis we have not attempted to model more
complicated redshift dependences of bias, the presence of abrupt degradations
in redshift etc. Such an analysis is beyond the scope of this work, but can be
done using similar tools once the details about the performance of a given 
 photometric method are known.

\section{Results -- Multiplicative error}
\label{sec:mult}

As explained in \S~\ref{sec:mult_define}, we adopt the multiplicative error of the form

\begin{equation}
\hat{P}^{\kappa}_{ij} (\ell)
=    P^{\kappa}_{ij} (\ell) \times\left [1+f_i+f_j\right ],
\end{equation}

\noindent where $\hat{P}^{\kappa}_{ij} (\ell)$ and $P^{\kappa}_{ij} (\ell)$ are
estimated and true convergence power respectively. In other words, the
multiplicative error we consider can be thought of as an irreducible but
perfectly coherent bias in the calibration of shear in any given redshift bin.

\begin{figure}[!t]
\psfig{file=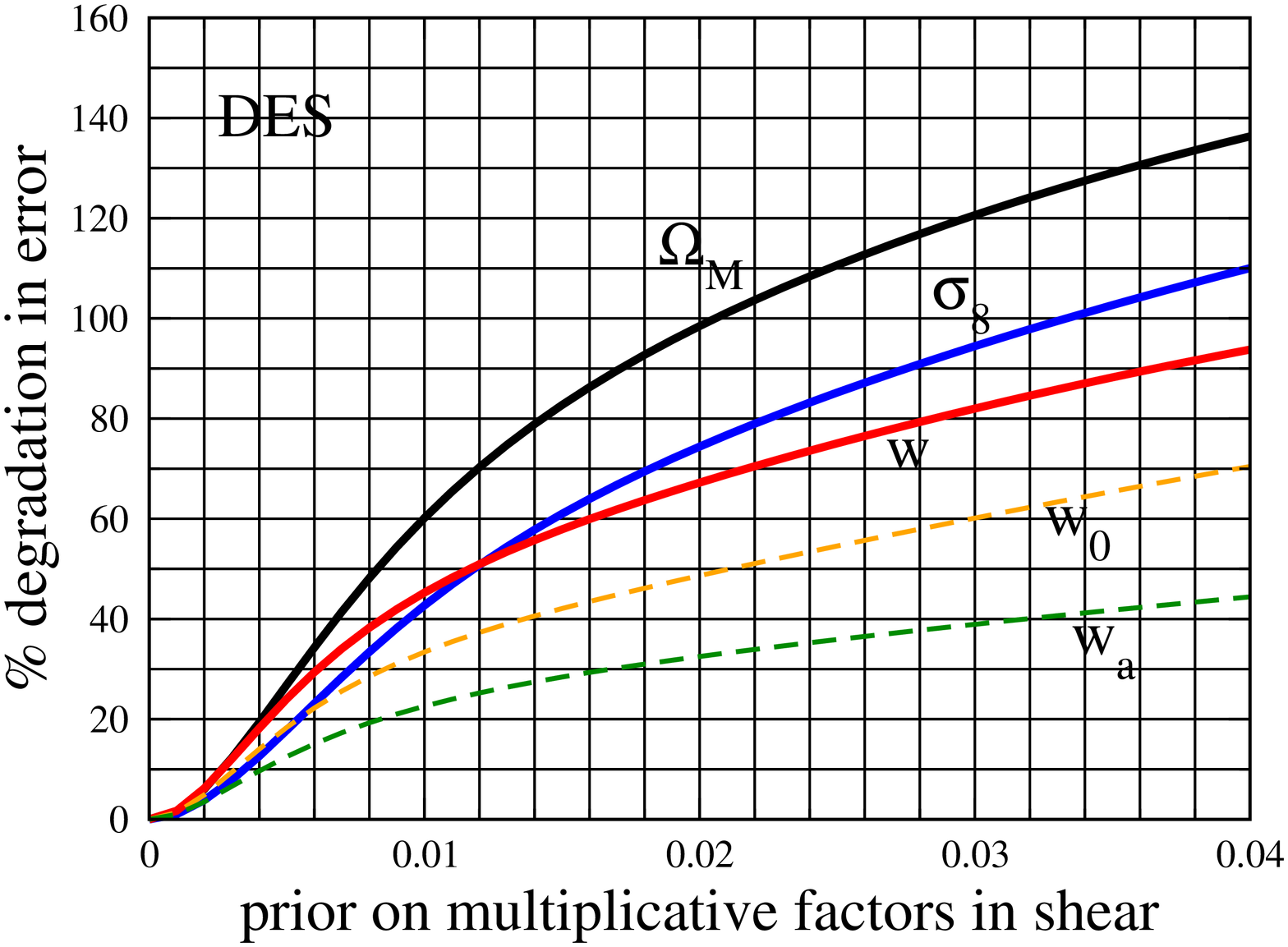, width=2.8in, height=3.5in, angle=-90}\hspace{-0.2cm}
\psfig{file=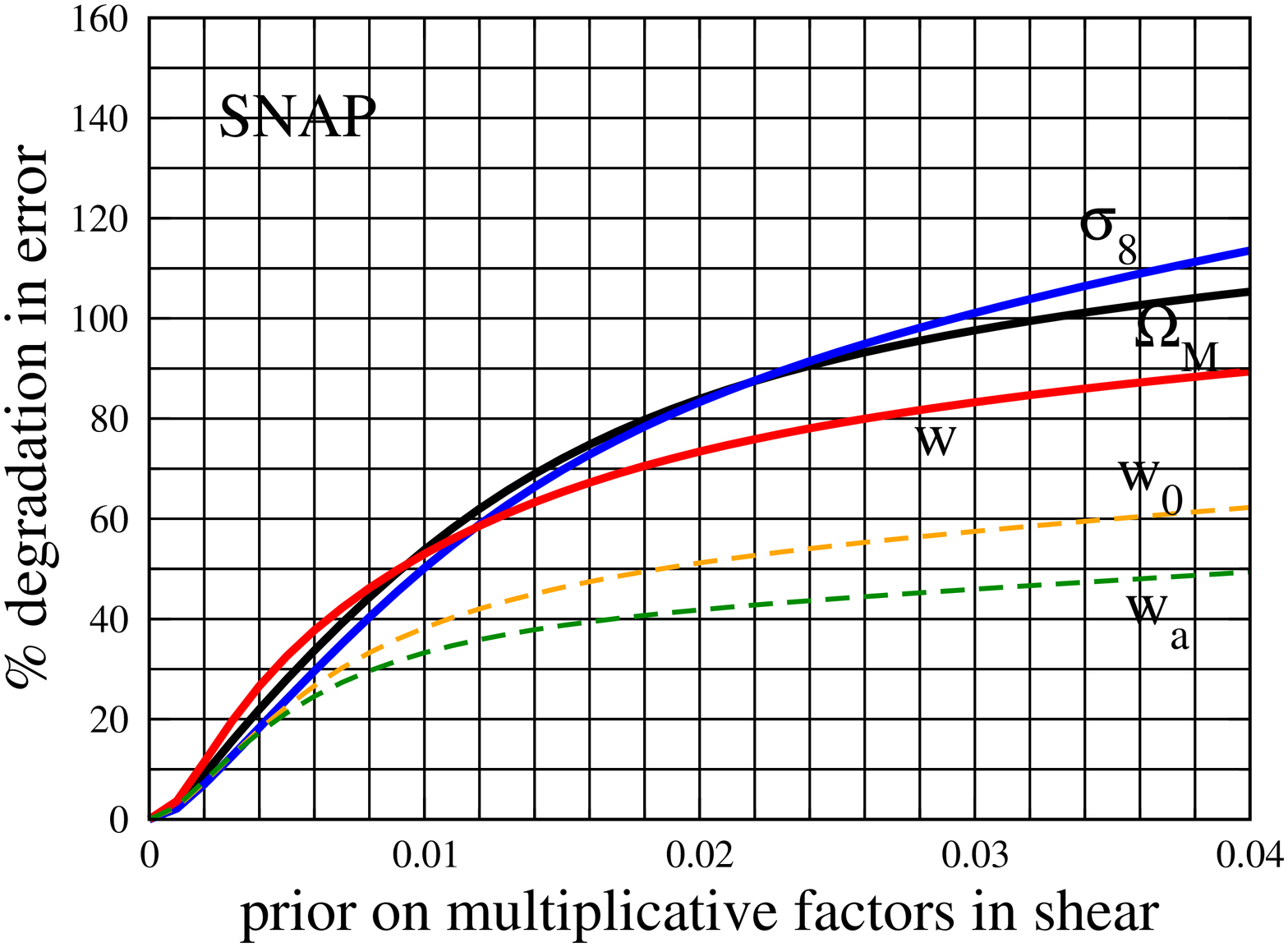, width=2.8in, height=3.5in, angle=-90}\\[-0.1cm]
\psfig{file=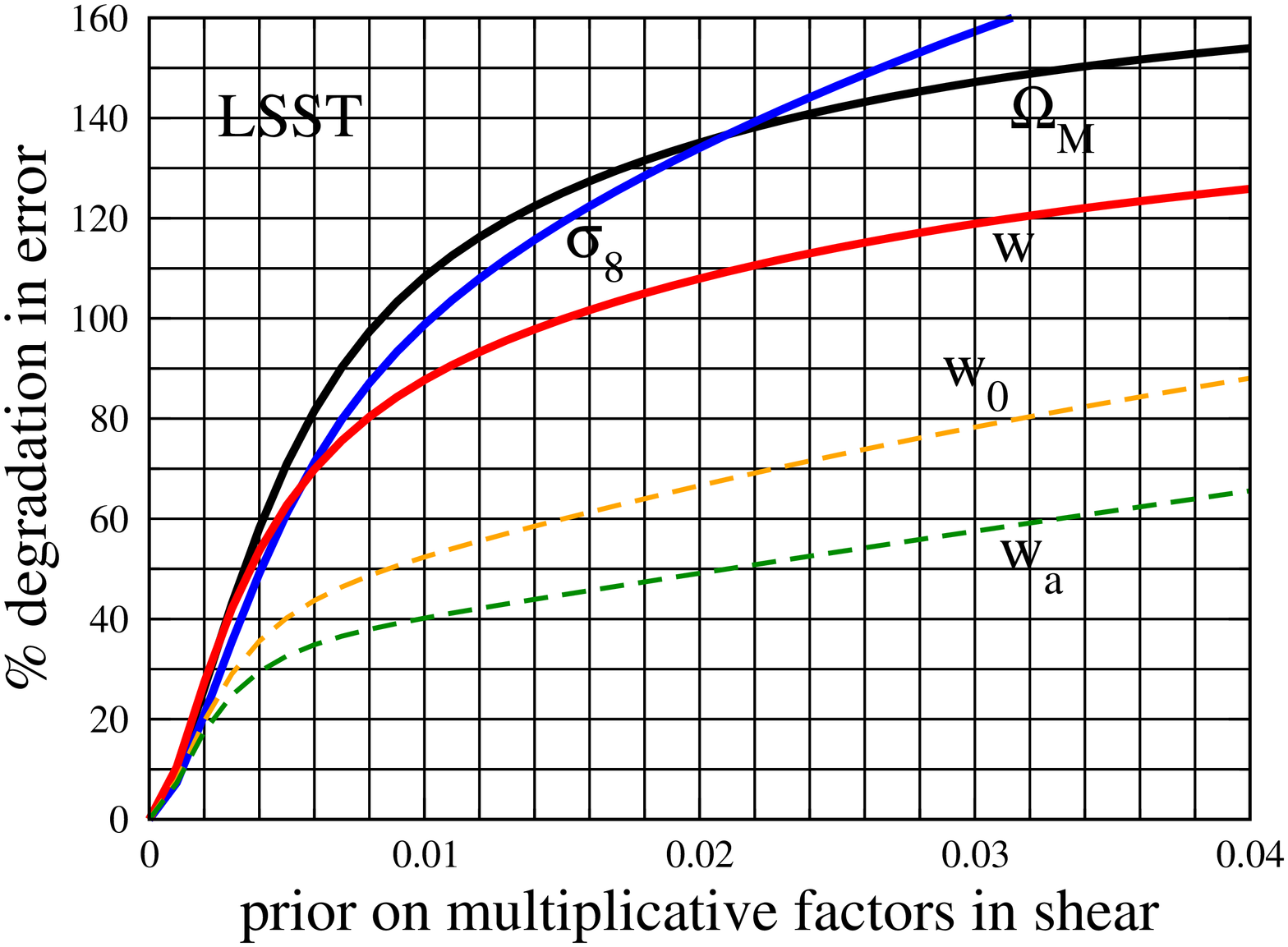, width=2.8in, height=3.5in, angle=-90} 
\caption{Degradation in marginalized errors in $\Omega_M$, $\sigma_8$ and
$w={\rm const}$, as well as $w_0$ and $w_a$, as a function of our prior
knowledge of the shear multiplicative factors. We give equal prior to
multiplicative factors in all redshift bins, and show results for the DES, SNAP
and LSST.  For example, existence of the multiplicative error of
$0.01\,(\fsky/f_{\rm sky, fid})^{-1/2}$ (or 1\% in shear for the fiducial sky
coverages) in each redshift bin leads to 50\% increase in error bars on
$\Omega_M$, $\sigma_8$ and $w$ for the DES and SNAP, and about a 100\%
degradation for LSST.}
\label{fig:mult}
\end{figure}

First, note that a tomographic measurement with $B$ redshift bins can determine
at most $B$ multiplicative parameters $f_i$ (this is true either if they are
just stepwise excursions as assumed here or coefficients of Chebyshev
polynomials described in the Appendix). For, if we had more than $B$
multiplicative parameters, at least one bin would have two or more parameters,
and there would be an infinite degeneracy between them. In contrast, 
the survey can in principle determine a much larger number of cosmological
parameters since they enter the observables in a more complicated way.  We
choose to use exactly $N_{\rm mult}=B$ multiplicative parameters (so $N_{\rm
mult}=10$ for SNAP and LSST and $N_{\rm mult}=7$ for the DES) since we would expect
the shear calibration to be different within each redshift bin. 

Figure \ref{fig:mult} shows the degradation in error in measuring three
cosmological parameters as a function of our prior knowledge of the
multiplicative factors; we give equal prior to all multiplicative factors.  For
the DES and SNAP, control of multiplicative error of $0.01\,(\fsky/f_{\rm sky,
fid})^{-1/2}$ (or 1\% in shear for the fiducial sky coverages) leads to a 50\%
increase in the cosmological error bars, and is therefore about the largest
error tolerable. For LSST, the requirement is about a factor of two more
stringent.

In general, it is interesting to consider whether the weak lensing survey can
'self-calibrate'', i.e.\ determine both the cosmological and the nuisance
parameters concurrently. This is partly motivated by the self-calibration of
cluster count surveys, where it has been shown that one can determine the
cosmological parameters and the evolution of the mass-observable relation --
provided that the latter takes a relatively simple deterministic form (e.g.\
Levine, Schulz \& White 2002, Majumdar \& Mohr 2003, Hu 2003b, Lima \& Hu 2004).
Figure ~\ref{fig:mult} shows that the fiducial DES and SNAP surveys 
can self-calibrate with about a 100\% degradation in cosmological parameter
errors (and LSST with about a 150\% degradation). While doubling the error in
cosmological parameters is a somewhat steep price to pay, it is very
encouraging that all surveys enter a self-calibrating regime where only the
higher-order moments of the error contribute to the total error budget.
In \S~\ref{sec:bispec} we show that the inclusion of bispectrum information
can significantly improve the self-calibration regime.

\section {Results: Additive error}
\label{sec:additive}

As explained in \S~\ref{sec:add_define}, we adopt the additive error of the form

\begin{equation}
P_{{\rm add}, ij}^{\kappa}(\ell)=\rho\, b_i b_j
\left ({\ell\over \ell_*}\right )^{\alpha},
\end{equation}

\noindent which adds that amount of noise to the convergence power spectra.
The coefficient $\rho $ is always $1$ for $i=j$, and its (fixed) value for $i\neq
j$ controls how much additive error leaks into the cross power spectra. We
weigh the fiducial value of $b_i$ by the inverse square of the average galaxy
size in $i$th redshift bin (or, by the square of the angular diameter distance
to the $i$th bin)\footnote{For $b_i=0$, the Fisher derivatives with
respect to $b_i$ are zero and no information about these parameters can
formally be extracted.}.

\begin{figure}
\psfig{file=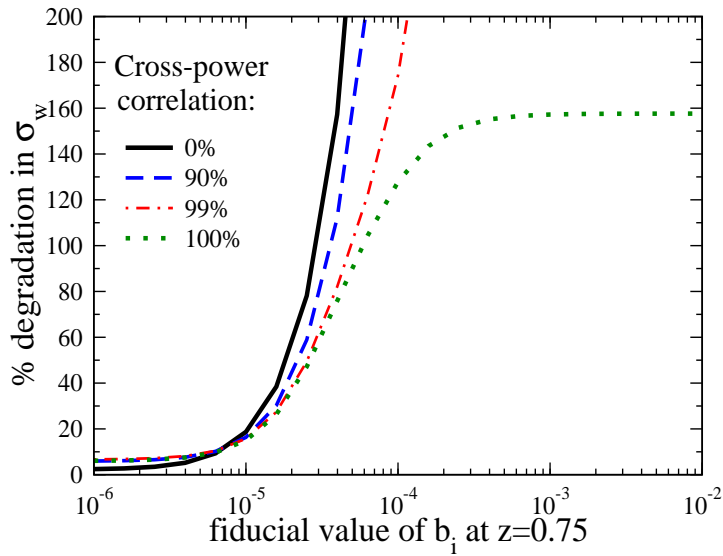, width=3.2in, height=4in, angle=-90}
\caption{Degradation in the cosmological parameter accuracies as a function of
  the {\it fiducial value} of the additive shear errors $b_i$, assuming no
  prior on the $b_i$. We show results for SNAP and for several values of the
  correlation coefficients between different bins, $\rho$, and
  marginalizing over the spatial power law exponent $\alpha$ which has fiducial value $\alpha=0$.
  The results are insensitive to various details, as discussed in the
  text, and the only exception is the possibility that the additive effect on
  the cross-power spectra is negligibly suppressed (i.e.  that the cross-power
  correlation is close to 100\%).  We conclude that the mean additive shear
  will need to be known to about $\sim 2\times 10^{-5}$, corresponding
  to shear variance of $\sim 10^{-4}$ on scales of $\sim 10$ arcmin.  }
\label{fig:add_error}
\end{figure}

Figure \ref{fig:add_error} shows the degradations in the equation of state $w$
as a function of the fiducial $b_i$ at redshift $z=0.75$ where, very roughly,
most galaxies are found (recall, the other $b_i$ are equal to this value modulo
order-unity weighting by the square of the angular diameter distance to their
corresponding redshift).  
%In the left panel of Fig.~\ref{fig:add_error} 
The solid line in the figure shows results for our fiducial SNAP survey, 
%two values of the power law $\alpha$, 
assuming no contribution to the cross-power spectra (i.e.\
$\rho=\delta_{ij}$). The coefficients $b_i$ need to be controlled to
$\sim 2\times 10^{-5}$, corresponding to shear variance of 
$\ell(\ell+1)P_{{\rm add}, ij}^{\kappa}(\ell)/(2\pi)\sim 10^{-4}$ on
scales of $\sim 10$ arcmin ($\ell\sim 1000$) where most constraining power of weak lensing resides.
The observed degradation in cosmological parameters is clearly due to the increased
sample variance that the additional power puts onto the measurement of the power spectrum.
When $\rho=1$, this sample variance is confined to a single mode of the shear
covariance matrix, so that the maximum damage is limited and we observe the
self-calibration plateau in Fig.~\ref{fig:add_error}.\footnote{One could in principle also
have a worst-case limit with additive errors whose functional form makes them
strongly degenerate with the effect of varying the cosmological parameters, where
the accuracy in cosmological constraints degrades as soon as the additive
power becomes comparable to the {\it measurement uncertainties} in the power spectrum
(and not the power spectrum itself, as above).}

We have tried varying a number of other details:

\begin{itemize}

\item Using different values of the coefficient $\rho$ for $i\neq j$ in the
range $0\leq \rho < 1$.

\item Changing the fiducial value of the exponent $\alpha$ from $0$
  to $+3$ or $-3$.

\item Adding a 10\% prior to the $b_i$ (rather than no prior).

\item Using the redshift independent fiducial values of the coefficients $b_i$.

\item Considering the degradation in the other cosmological parameters.

\end{itemize}

Interestingly, we find that the overall requirements are very weakly dependent
on any of the above variations, and the requirements almost always look roughly
like those in Figure \ref{fig:add_error}. In particular, the results are
essentially insensitive to reasonable priors of any of the nuisance parameters,
since the $b_i$ can be determined internally to an accuracy
much better than $b_i$ for all but the smallest ($\lesssim 10^{-7}$) fiducial values of
these parameters.  Therefore the dominant effect by far is the
fiducial value of $b_i$ and the increased sample variance  that it introduces.

The only interesting exception to this insensitivity is allowing the $i\neq j$
value of the coefficients $\rho$ to be very close to unity. In the extreme
case when $\rho =1$ for $i\neq j$, the degradation asymptotes to $\sim 150\%$
even with very large $b_i$ since in that case the additive errors add a huge
contribution to {\it all} power spectra, and one can mathematically show that
the resulting errors in cosmological parameters do not change more than $\sim 100\%$
irrespective of the systematic error.  In practice, however,
$\rho\gtrsim 0.99$ for $i\neq j$ is needed to see an appreciable difference (see the
other curves in Fig.~\ref{fig:add_error}), and it is reasonable to believe that
the actual errors in the cross-power spectra will be suppressed by much more
than a percent, resulting in the degradations as shown by the solid curve in
Fig.~\ref{fig:add_error}.

While our model for the additive errors is admittedly crude, it is very
difficult to parameterize these errors more accurately without end-to-end
simulations that describe various systematic effects and estimate their
contribution to the additive errors (such simulations are now being planned or
carried out by several research groups). Moreover, additive errors cannot be
self-calibrated unless we can identify a functional dependence, 
which is distinct from the cosmological dependence of the power spectrum, that it must
have. Finally, we note that space-based surveys, such as SNAP, are expected to
have a more accurate characterization of the additive error, primarily from the
absence of atmospheric effects in the characterization of the PSF.

\section{Systematics with the Bispectrum}
\label{sec:bispec}

We now extend the calculations to the bispectrum of weak gravitational lensing.
The bispectrum is the Fourier counterpart of the three-point correlation
function, and it describes the non-Gaussianity of mass distribution in the
large-scale structure that is induced by gravitational instability from the
primordial Gaussian perturbations that the simplest inflationary models predict.
The redshift evolution and configuration dependence of the mass bispectrum can
be accurately predicted using a suite of high-resolution $N$-body simulations
(see e.g.\ Bernardeau et al.\ 2002 for a comprehensive review).  The bispectrum
of lensing shear arises from the line-of-sight integration of products of the
mass bispectrum and the lensing geometrical factor (see Eq.~(18) in Takada \&
Jain 2004).  Following Jain \& Seljak (1997), we roughly estimate
how the lensing power spectrum and bispectrum scale with cosmological
parameters by perturbing around the fiducial $\Lambda$CDM model:

\begin{figure}
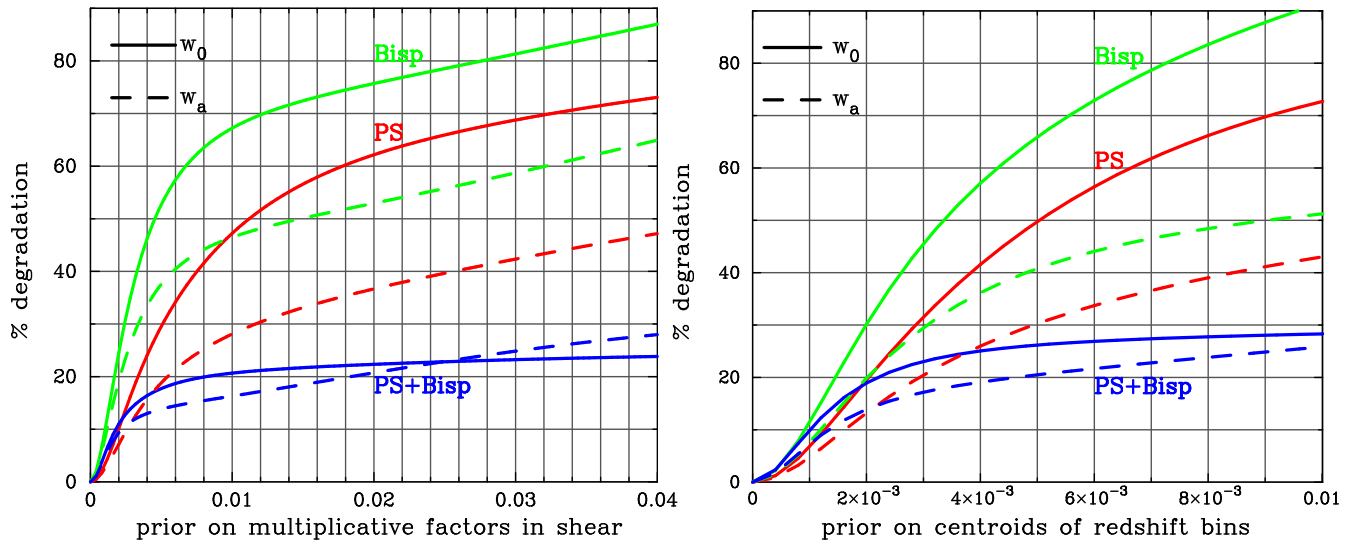

\psfig{file=zeta_v3.ps, width=2.8in, height=3.5in, angle=-90}\hspace{-0.2cm}
\psfig{file=photoz_v3.ps,width=2.8in,height=3.5in, angle=-90}
\caption{Degradation in $w_0$ and $w_a$ accuracies expected for the SNAP as a
function of our prior knowledge of the multiplicative error in shear (left
panel) and redshift centroids (right panel).  The priors on the multiplicative
or redshift parameters are shown on the x-axis. While the power spectrum (PS) and bispectrum (BS)
individually enter a self-calibration regime with $\sim 100\%$ degradation in
cosmological parameter errors, combining the two leads to self-calibration with
only a $20$-$30$\% degradation. This improvement in the self-calibration limit
is in addition to the already smaller no-systematic error bars with PS+BS as
compared to either PS or BS alone. }
\label{fig:bispec_wa}
\end{figure}

\begin{equation}
P^\kappa\propto \Omega_{\rm
DE}^{-3.5}
\sigma_8^{2.9}z_{\rm s}^{1.6}|w|^{0.31},  
\hspace{2em}B^\kappa\propto \Omega_{\rm
DE}^{-6.1} \sigma_8^{5.9}z_{\rm s}^{1.6}|w|^{0.19},
\label{eqn:expand}
\end{equation}
where we have considered multipole mode of $l=1000$, equilateral triangle
configurations in multipole space, redshift of all source galaxies $z_{\rm
s}=1$ (no tomography), and constant equation of state parameter $w$. We adopt
the model described in Takada \& Jain (2004) to compute the lensing bispectrum.
Equation (\ref{eqn:expand}) illustrates that the power spectrum and the
bispectrum depend upon specific combinations of cosmological parameters. For
example, the power spectrum (and more generally any two-point statistics of
choice) depends on $\Omega_{\rm DE}$ and $\sigma_8$ through the combination
$\Omega_{\rm DE}^{-1.2}\sigma_8$ (or equivalently
$\Omega_{M}^{0.5}\sigma_8$). Furthermore, the cosmological parameter
dependences are strongly degenerate with source redshift ($z_{\rm s}$) unless
accurate redshift information is available.  Importantly, the power spectrum
and bispectrum have substantially {\it different} dependences on the
parameters, suggesting that combining the two can be a powerful way of breaking
the parameter degeneracies. For example, it is well known that a combination of
$B/P^2$, motivated from the hierarchical clustering ansatz, depends mainly on
$\Omega_{\rm DE}$, with rather weak dependence on $\sigma_8$, so that roughly
$B/P^2\propto \Omega_{\rm DE}^{0.9} \sigma_8^{-0.1}$ (Bernardeau et al. 1997,
Hui 1999, Takada \& Jain 2004).  As another example, the bispectrum amplitude
increases with the mean source redshift $z_{\rm s}$ more slowly than $P^2$
because the non-Gaussianity of structure formation becomes suppressed at higher
redshifts.  It is therefore clear that photometric redshift errors of source
galaxies will affect the power spectrum (PS) and bispectrum (BS)
differently. Likewise, it is natural to expect that other systematics that we
have considered affect the power spectrum and the bispectrum in a different
way.

For the PS+BS systematic analysis, we consider the redshift and multiplicative
errors but not the additives. For the redshift errors we adopt the parameterization of 
redshift bin centroids, now applied to tomographic bins of both PS and BS.
The multiplicative error is modeled similarly as in \S
\ref{sec:mult_define}: the observed bispectrum with tomographic redshift bins $i$,
$j$ and $k$, $\hat{B}_{ijk}(l_1,l_2,l_3)$, is related to the true bispectrum 
$B_{ijk}(l_1,l_2,l_3)$ via
\begin{equation}
\hat{B}_{ijk}(l_1,l_2,l_3)=B_{ijk}(l_1,l_2,l_3)\times [1+f_i+f_j+f_k].
\label{eq:define_mult_BS}
\end{equation}

\begin{figure}
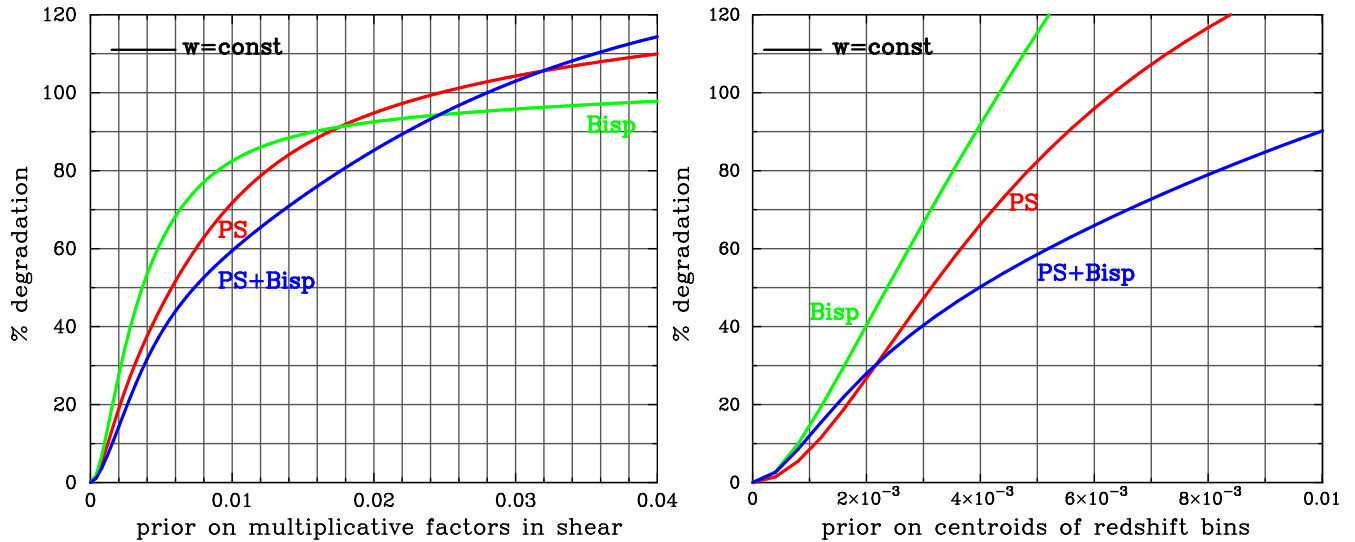

\psfig{file=zeta_wafixed_v2.ps,width=2.8in,height=3.5in,angle=-90}\hspace{-0.2cm}
\psfig{file=photoz_wafixed_v2.ps,width=2.8in, height=3.5in, angle=-90}
\caption{Degradation in $w={\rm const}$ for the multiplicative errors (left
 panel) and redshift errors (right panel).  Note that the low self-calibration
 asymptote seen in Fig.~\ref{fig:bispec_wa} is not seen any more, and even the
 cases when PS+BS are combined lead to appreciable degradations.  We conclude
 that $w={\rm const}$ (or more generally any accurately measured combination of
 $w(z)$) does not benefit from self-calibration much, but $w_0$ and $w_a$
 individually do as shown in Fig.~\ref{fig:bispec_wa}.  }
\label{fig:bispec_w}
\end{figure}

Based on the considerations above, we study how combining the power spectrum
with the bispectrum allows us to break parameter degeneracies not only in
cosmological parameter space but also those induced by the presence of the
nuisance systematic parameters. We include all triangle configurations, and
compute the bispectra constructed using 5 redshift bins up to $l_{\rm
max}=3000$. Note that, for a given triangle configuration, there are $5^3$
tomographic bispectra; the large amount of time necessary to compute them is
the reason why we use 5 tomographic bins instead of the original
7-10.\footnote{For the same reason, the PS curves in Figs.~\ref{fig:bispec_wa}
and \ref{fig:bispec_w} do not exactly match those in Figs.~\ref{fig:z} and
\ref{fig:mult}.} Note too that we have assumed that the PS and BS are
uncorrelated and simply added their Fisher matrices when combining them. While
they are strictly uncorrelated in linear theory, nonlinear structure formation
will introduce the correlation between the two, and the full information
content will be smaller than our estimate\footnote{We thank Martin White for
drawing our attention to this issue.}. Correlation between the PS and BS has not been
accurately estimated to date, and such a project is well beyond the scope of our
paper.  However, our main emphasis here is not to accurately estimate the
resulting cosmological error bars but rather to study the overall effect of the
systematics when different weak lensing probes are combined. We expect the qualitative
trends (discussed below) to be unchanged in cases when we add information from
cross-correlation cosmography or cluster counts to the PS.

Figure \ref{fig:bispec_wa} shows how the multiplicative errors (left panel)
and redshift centroid errors (right panel) degrade constraints on $w_0$
and $w_a$ expected for SNAP.  Each panel shows the degradation in the parameter
measurement due to the power spectrum, bispectrum, and the two combined.  The
most remarkable fact seen in Fig. \ref{fig:bispec_wa} is that the {\it
degradation} in the parameter accuracies is smaller with PS and BS combined as
opposed to either one separately. This is because dependences of BS on
cosmological parameters as well as the model systematics are complementary to
those from PS.  Therefore, combining the PS and BS has a very beneficial effect
of protecting against the systematics by more than a factor of two than either
statistic alone!

However, the drastic improvement in self-calibration does not hold up for
parameters that are more accurately measured, such as $w={\rm const}$.  Figure
\ref{fig:bispec_w} shows the degradation in the $w={\rm const}$ case for the
multiplicative errors (left panel) and redshift errors (right panel).  It is
clear that the low self-calibration asymptote seen in Fig.~\ref{fig:bispec_wa}
is no longer present, and even the cases when PS and BS are combined lead to
appreciable degradations.  Therefore, $w={\rm const}$ does not benefit from
self-calibration as much as $w_0$ and $w_a$.  More generally, self-calibration
with PS+BS works much better when applied to poorly determined combinations of
cosmological parameters than to the accurately determined ones; we have
explicitly checked this by diagonalizing the full Fisher matrix and finding the
degradations in all of its eigenvectors.

\section {Combining the systematic errors}\label{sec:discussion}

In Table \ref{tab:errors} we now show the principal cosmological parameter with
their fiducial values, and their errors for the PS only and PS+BS cases.  In
each case we show the error without any systematics, and the error after adding
sample systematic errors.  For the systematics, we assume redshift biases
(described by Chebyshev polynomials) with priors of 0.0005, {\it together with}
the multiplicative errors of 0.005. The errors have been marginalized over the
other cosmological parameters, and the systematic-case errors have further been
marginalized over the 30 redshift and 10 multiplicative nuisance parameters
with the aforementioned priors. While the systematics used are a guess -- and
they are likely to remain uncertain well into the planning phase of each survey
-- our goal here is to see how the errors degrade. While the errors in the
strongest fiducial surveys get degraded the most, we find that strongest
surveys remain in that position even after adding the systematics. For example,
LSST's fiducial error on $w$ was a factor of two better than SNAP's before
adding the systematics, and is 50\% better after adding them. Note, however,
that we have added {\it equal} systematic errors to all three surveys in this
example; in reality, a space-based survey like SNAP is expected to have a
better control of the systematics than the ground based surveys 
(see e.g.\  Rhodes et al.\ 2004).

\begin{figure}
\psfig{file=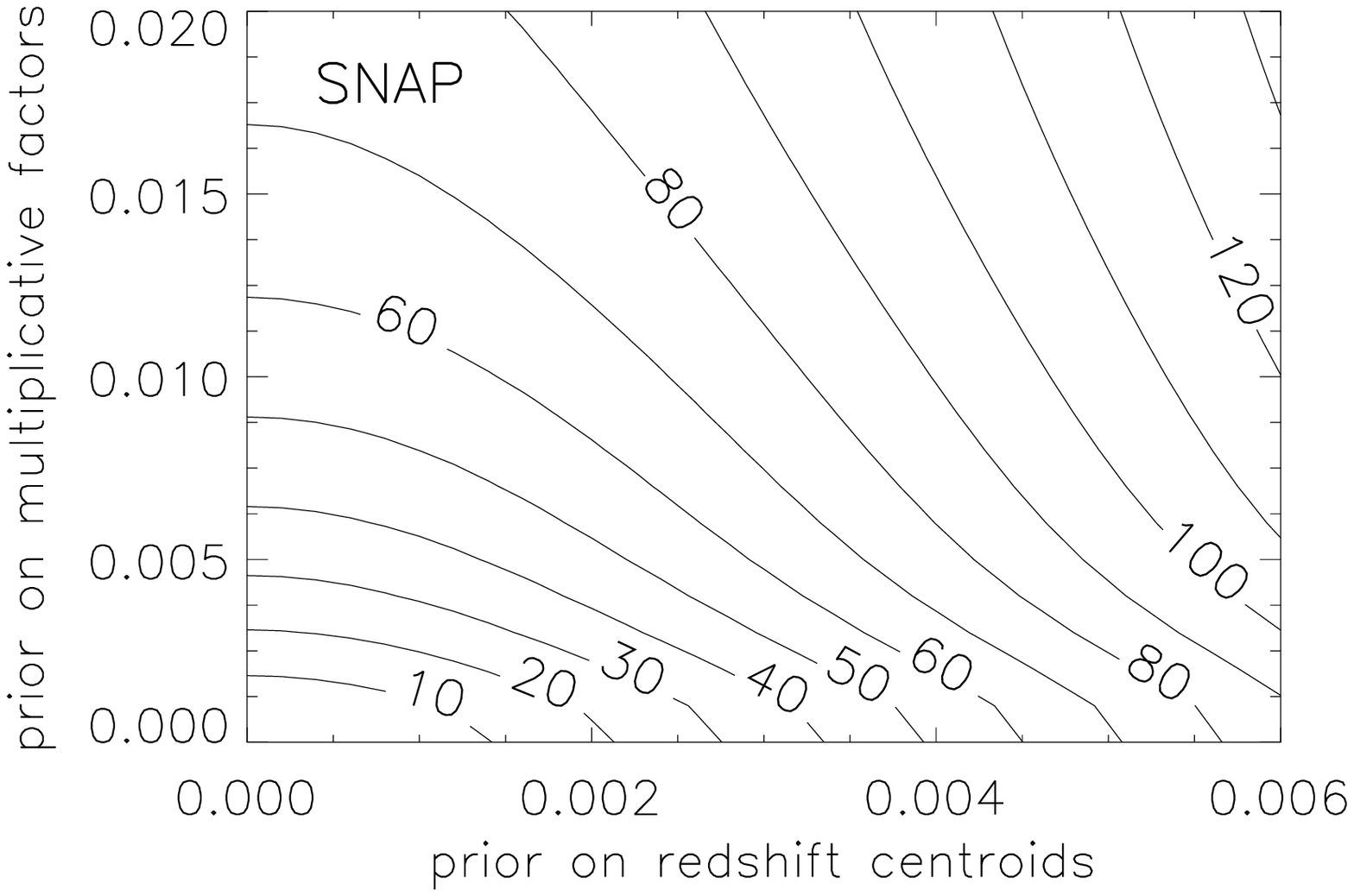,width=3.5in,height=3in}\hspace{-0.2cm}
\psfig{file=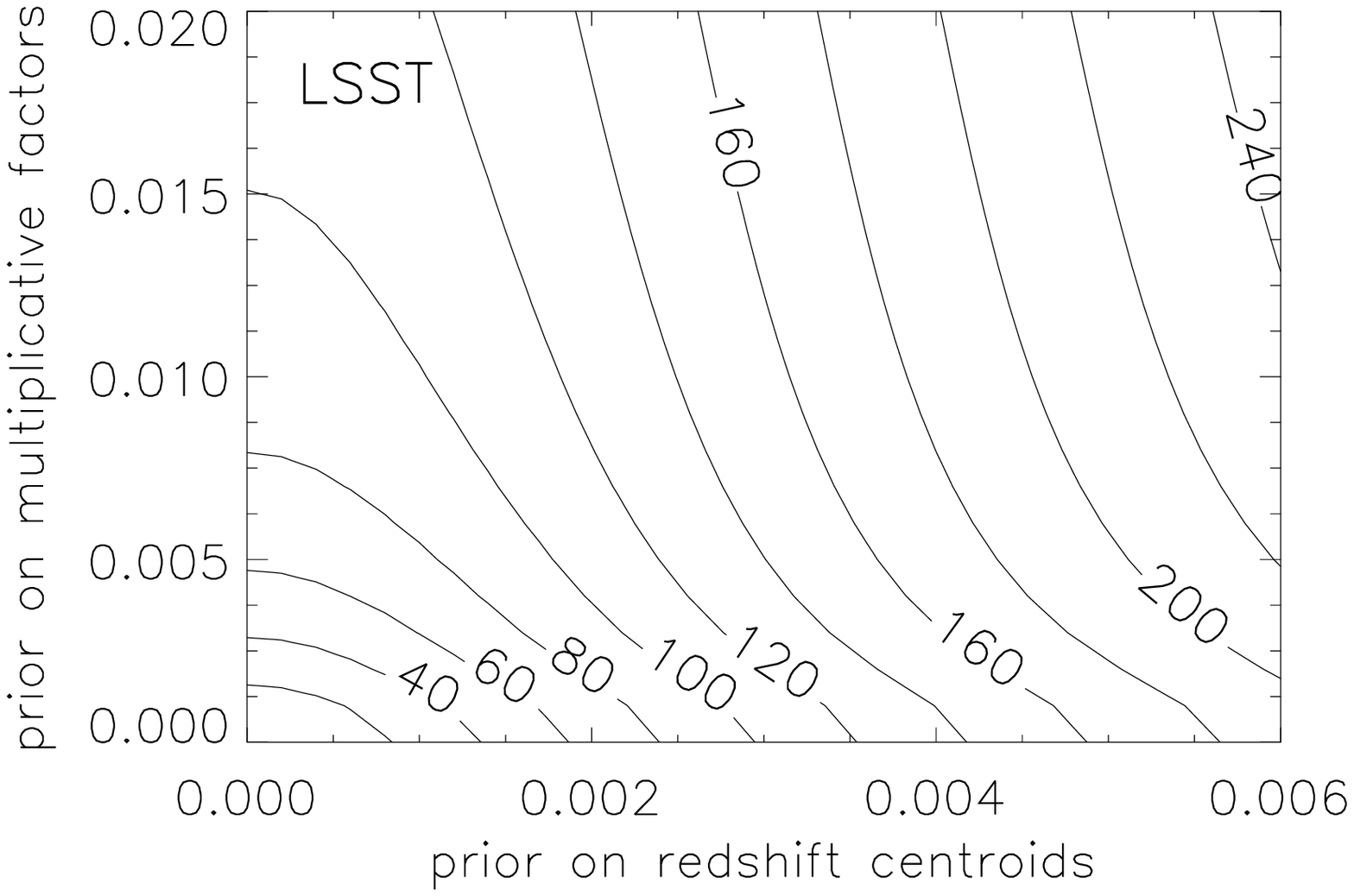,width=3.5in, height=3in}
\caption{Contours of constant degradation in equation of state of dark energy
$w$ (for the PS case only) with both redshift and multiplicative errors
included and for the SNAP fiducial survey (left panel) and LSST (right
panel). As before, the degradations are unchanged for a different survey area
if we let both errors scale as $\fsky^{-1/2}$. Note that the contours become
progressively more vertical at larger values of the redshift error -- therefore,
the results are weakly affected by the multiplicative errors once the redshift
errors become substantial. 
}
\label{fig:cont}
\end{figure}

In Fig.~\ref{fig:cont} we show the contours of constant degradation in
equation of state of dark energy $w$ (for the PS case only) with both redshift
and multiplicative errors included. The x and y-axis intercepts of degradation
values in this Figure correspond to the case of redshift centroid errors only
(as in Fig.~\ref{fig:z}) and multiplicative errors only (as in
Fig.~\ref{fig:mult}) respectively, but the rest of the plane allows for the simultaneous
presence of both types of error.  Note that the contours become progressively
more vertical at larger values of the redshift error -- therefore, the results
are weakly affected by the multiplicative errors once the redshift errors
become substantial. However, we would ideally like to be in the regime where
the degradation is smaller than $\sim 50\%$, and in that case both types of error (as well as
the additives) need to be controlled to a correspondingly good accuracy as discussed in
\S~\ref{sec:redshift_centroids} and \S~\ref{sec:mult}.

\begin{table}
\begin{tabular}{||c||c||c|c||c|c|||c|c||}
\hline\hline
\rule[-3mm]{0mm}{8mm} Parameter & Fiducial value 
& DES  (PS) 
& DES  (PS+BS) 
& SNAP (PS) 
& SNAP  (PS+BS) 
& LSST (PS) 
& LSST (PS+BS) 
 \\\hline
\rule[-3mm]{0mm}{8mm} $\Omega_{\rm M}$ & 0.3 &
0.008/0.011 & 0.006/0.009 & 0.008/0.011 & 0.004/0.006 & 0.003/0.007 & 0.002/0.005
\\ \hline 
\rule[-3mm]{0mm}{8mm} $w$ & -1.0 & 
0.092/0.120  & 0.035/0.060  & 0.058/0.081 & 0.027/0.036 & 0.029/0.053 & 0.010/0.023
\\ \hline 
\rule[-3mm]{0mm}{8mm} $\sigma_8$  & 0.9 & 
0.010/0.012 & 0.006/0.008 & 0.008/0.011 & 0.005/0.008 & 0.004/0.006 & 0.003/0.004
\\ \hline\hline 
\rule[-3mm]{0mm}{8mm} $w_0$ & -1.0 &
0.33/0.40   & 0.18/0.20   & 0.28/0.35   & 0.09/0.12   & 0.13/0.20   & 0.06/0.07
\\ \hline
\rule[-3mm]{0mm}{8mm} $w_a$ & 0  &
1.41/1.64   & 0.82/0.92   & 0.96/1.20   & 0.35/0.44   & 0.49/0.69   & 0.25/0.27
\\ \hline\hline 
\end{tabular}
\caption{Cosmological parameter errors without the systematics (numbers
preceding the slash in each box) and with sample systematics (numbers following
the slash).  For the systematics case we assumed redshift biases (described by
Chebyshev polynomials) of 0.0005, {\it together with} the multiplicative errors
of 0.005. The errors are shown for the power spectrum tomography only (PS), and for
the power spectrum and the bispectrum (PS+BS). Errors in other cosmological parameters
($\Omega_M h^2$, $\Omega_B h^2$, $n$) are not shown.}
\label{tab:errors}
\end{table}

\section {Conclusions}\label{sec:conclude}

We considered three generic types of systematic errors that can affect a weak
lensing survey: measurements of source galaxy redshifts, and multiplicative
and additive errors in the measurements of shear. 
We considered three representative future wide-field  surveys (DES,
SNAP and LSST) and used weak lensing tomography with either power 
spectrum alone, or power spectrum and bispectrum combined.

The most important (and difficult) part is to parameterize the systematic
errors. We are solely interested in the part of the systematics that has not
been corrected for in the data analysis, as that part can lead to 
errors in the estimated cosmological parameters. 
For the redshift error we adopt two alternative
parameterizations 
that should both be useful in calibrating photometric redshift methods. 
Multiplicative errors in shear measurement are described by
one parameter for the error in shear in
each given redshift bin.  The additive errors are most difficult to parameterize,
as their redshift dependence and spatial coherence both need to be 
specified; our treatment of the additive errors, 
while robust with respect to various details of the model, is
only a first step and can be improved upon with further simulation of realistic
systematics.

In general, higher fiducial accuracy in the cosmological parameters leads to
more stringent requirements on the systematics. Therefore, LSST typically has
the most stringent requirements, followed by SNAP and then the
DES (at their fiducial $f_{\rm sky}$). 
We note that the {\it accurately determined} combinations of dark
energy parameters typically are more sensitive to systematics than 
the {\it poorly determined} combinations.  At this point one
could ask, which is the important quantity: the individual parameters (say,
$w_0$ and $w_a$), or their linear combination that is well measured by the
survey? We argue that both are needed: the former to understand the behavior of
dark energy at any given redshift, and the latter to maximize the constraining
power of weak lensing when combining it with other cosmological probes.

For a SNAP-type survey with $\sim 100$ gal/arcmin$^2$, we find that the
centroids of redshift bins (of width $\Delta z=0.3$) need to be known to about
0.003$\,(\fsky/f_{1000})^{-1/2}$ in order not to lead to parameter degradations
larger than $\sim 50\%$. For the LSST-type survey the requirements
are about a factor of two more stringent, i.e.\
0.0015$\,(\fsky/f_{15000})^{-1/2}$. These numbers correspond to controlling
each Chebyshev mode of smooth variations in $z_p-z_s$ to about
0.001$\,(\fsky/f_{1000})^{-1/2}$ and 0.0005$\,(\fsky/f_{15000})^{-1/2}$
respectively. These requirements would easily be satisfied by planned surveys
if the bias were due to residual statistical errors, since these surveys will
have well over a million galaxies per redshift bin.
But it remains a challenge to control the {\it systematic}
biases to this level, presumably by using the spectroscopic training sets. 

The multiplicative errors need to be determined to about $0.01\,(\fsky/f_{\rm
sky, fid})^{-1/2}$ (or 1\% of average shear in a given tomographic redshift
bin, for the fiducial sky coverages) for the DES and SNAP. The requirements are
about a factor of two more stringent for the LSST. The actual multiplicative
error will depend on the galaxy size and shape, and might have spatial
dependence as well. The numbers we quote here refer to the post-correction
systematic error, averaged over all galaxies and directions in the
sky.

The additive errors in shear require more detailed modeling than the
multiplicatives, and in particular they require specifying its
two-point correlation
function (and the three-point function if we are to consider
the bispectrum measurements).  We constructed a simple model 
that includes the redshift and angular dependence of the additive power and a
coefficient that specifies the effect on the cross-power spectra relative to
that on the auto-spectra.  In most cases the additive error in each redshift
bin needs to be controlled to a few times $10^{-5}$,  which corresponds to shear variance of 
$\sim 10^{-4}$ on  scales of $\sim 10$ arcmin ($\ell\sim 1000$).
%or roughly 1\% or less of the shear. 
Note too that the additive error cannot be self-calibrated unless we
can identify a functional dependence that it must have. Our parameterization of
the additive errors is a first step, and improvements can be made once the
sources of the additive error are studied in more detail both from the data and
via ray-tracing simulations for given telescope designs.

While the systematic requirements are not stringent beyond what one can
reasonably hope to achieve with upcoming surveys, perhaps the most encouraging
aspect that we highlighted is the possibility of self-calibration of a part of
the systematics.  In this scheme, weak lensing data is used to concurrently
determine both the systematic and the cosmological parameters.  The effects of
the parameterized systematics can then be marginalized out without the need to
know their values (but at the expense of increasing the cosmological parameter
errors), leaving only the subtler systematic effects that were effectively
not taken into account with the assumed parameterization.  
We find that
power spectrum measurements can lead to self-calibration with $\sim 100\%$
error degradation in most cases. A promising result is  that 
combining the PS and BS
measurements leads to self-calibration with 20--30\%
degradation, at least for the more poorly-constrained combinations of
parameters.  Therefore, not only are the fiducial constraints of PS+BS better
than those with PS alone, but also the {\it degradations} relative to their
fiducial constraints are smaller in the PS+BS case, simply because the combined
PS+BS are more effective in breaking the degeneracies between the systematic
and cosmological parameters than the PS or the BS alone. However, we also found
that the self-calibration with PS+BS is not nearly as effective if we consider
$w={\rm const}$ (or the best-determined combination of $w_0$ and $w_a$). More
generally, the accurately measured principal component of $w(z)$ does not
benefit from self-calibration as much as its poorly measured components.
Finally, we only considered the constraints from the PS and BS, without using information from
cross-correlation cosmography or cluster counts.  Including the latter two 
methods, which we plan to do in the near future, could further improve the prospects 
for self-calibration of systematic errors.

\section*{Acknowledgments}
DH is supported by the NSF Astronomy and Astrophysics Postdoctoral Fellowship
under Grant No.\ 0401066. GMB acknowledges support from grant AST-0236702 from the
National Science Foundation, and Department of Energy grant
DOE-DE-FG02-95ER40893. We thank Carlos Cunha, Josh Frieman,  Mike Jarvis, Lloyd
Knox, Marcos Lima, Eric Linder, Erin Sheldon, Zhaoming Ma, Joe Mohr, Hiroaki
Oyaizu, Fritz Stabenau, Tony Tyson, Martin White, and especially Wayne Hu
for many useful conversations.

\appendix
\section{Chebyshev Polynomials}

We would like to parameterize the bias between the photometric and spectroscopic
redshifts, $\delta z \equiv z_p - z_s$, as a function of $z_s$. This function
is expected to be relatively smooth, and one promising way to parameterize it is
to use Chebyshev polynomials.  Chebyshev polynomials of the first kind
$T_i(x)$ ($i=0, 1, 2, \ldots$) are smooth functions, orthonormal in the
interval $x=[-1, 1]$, and take values from $-1$ to 1. The first two are
$T_0(x)=1$ and $T_1(x)=x$.

One can represent the redshift uncertainty in terms of the first  $M$ 
Chebyshev polynomials as

\begin{equation}
\delta z \equiv z_p - z_s = \sum_{i=1}^{M} \,g_i \,
T_i\left ({z_s-z_{\rm max}/2 \over z_{\rm max}/2}\right )
\end{equation}

\noindent where $z_{\rm max}$ is the maximum extent of the galaxy distribution
in redshift. For convenience, the fiducial values for the extra parameters are
taken to be $g_i=0$ ($i=0, 1, 2, \ldots, M-1$). Then the derivatives with
respect to the nuisance parameters can be computed via $d/dg_k =
[d/dz_p]\,[dz_p/dg_k]$.

\begin{figure}
\psfig{file=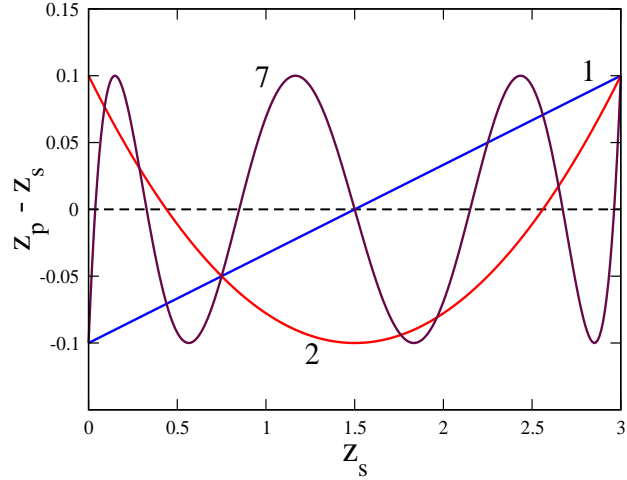, width=2.8in, height=3.5in, angle=-90}\\[0.1cm]
\caption{Select three modes (first, second and
seventh) of perturbation to the relation between the photometric and
spectroscopic redshift using the Chebyshev polynomials out to $z_{\rm max}=3$.}
\label{fig:deltaz_vs_zs}
\end{figure}

Figure \ref{fig:deltaz_vs_zs} shows the select three modes
(first, second and seventh) of perturbation to the relation between the
photometric and spectroscopic redshift.

\end{document}